\documentclass[12pt]{article}
\usepackage{amsmath}
\usepackage{graphicx}

\newsavebox{\SKpath}
\sbox{\SKpath}{\includegraphics[width=8.8cm]{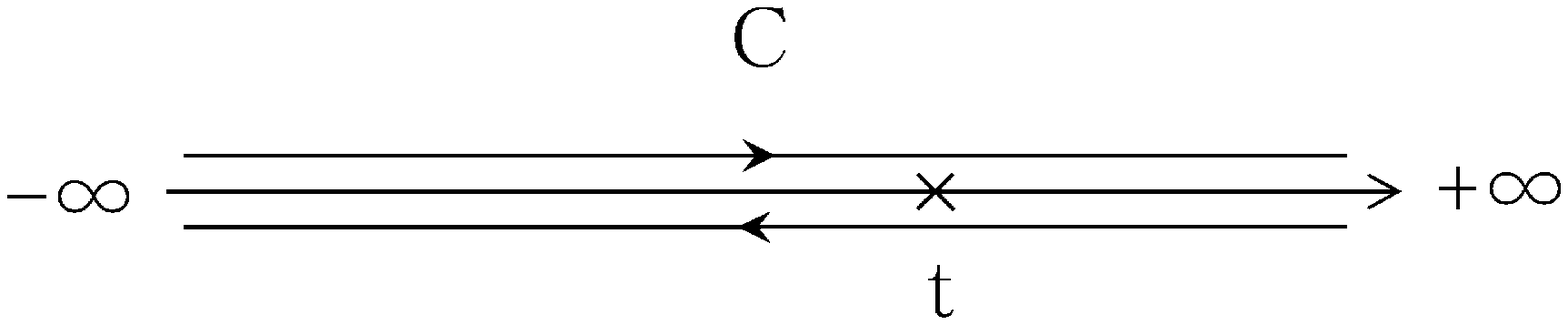}} 
\newlength{\SKpathl}
\def\papertitlepage{\baselineskip 3.5ex\thispagestyle{empty}}
\def\preprinumber#1#2{\hfill\begin{minipage}{4.2cm} #1
        \par\noindent #2 \end{minipage}}
\allowdisplaybreaks[3]

\begin{document}

\papertitlepage
\setcounter{page}{0}
\preprinumber{KEK-TH-1525}{}
\baselineskip 0.8cm
\vspace*{2.0cm}

\begin{center}
{\Large\bf Soft Gravitons Screen Couplings\\
 in de Sitter Space}
\end{center}

\begin{center}
Hiroyuki K{\sc itamoto}$^{1)}$\footnote{E-mail address: kitamoto@post.kek.jp} and
Yoshihisa K{\sc itazawa}$^{2),3)}$\footnote{E-mail address: kitazawa@post.kek.jp}\\
\vspace{5mm}
$^{1)}${\it Department of Physics and Astronomy\\
Seoul National University, 
Seoul 151-747, Korea}\\
$^{2)}${\it KEK Theory Center, 
Tsukuba, Ibaraki 305-0801, Japan}\\
$^{3)}${\it Department of Particle and Nuclear Physics\\
The Graduate University for Advanced Studies (Sokendai)\\ 
Tsukuba, Ibaraki 305-0801, Japan}
\end{center}

\vskip 5ex
\baselineskip = 2.5 ex

\begin{center}{\bf Abstract}\end{center}

The scale invariance of the quantum fluctuations in de Sitter space leads to the appearance of de Sitter symmetry breaking infra-red logarithms in the graviton propagator.
We investigate physical effects of soft gravitons on the local dynamics of matter fields well inside the cosmological horizon.
We show that the IR logarithms do not spoil Lorentz invariance in scalar and Dirac field theory.
The leading IR logarithms can be absorbed by a time dependent wave function renormalization factor in the both cases.
In the interacting field theory with $\lambda \phi^4$ and Yukawa interaction, we find that the couplings become time dependent with definite scaling exponents.
We argue that the relative scaling exponents of the couplings are gauge invariant and physical as we can use the evolution of a coupling as a physical time.

\vspace*{\fill}
\noindent
June 2013

\newpage
\section{Introduction}
\setcounter{equation}{0}

In de Sitter (dS) space, the degrees of freedom outside the cosmological horizon increase with cosmic evolution. 
This increase leads to the dS symmetry breaking term in the propagator of a massless and minimally coupled scalar field. 
It is a direct consequence of the scale invariant fluctuation spectrum \cite{Vilenkin1982,Linde1982,Starobinsky1982}. 
So in some field theoretic models in dS space, physical quantities may become time dependent through the propagator. 
The gravitational field on dS background is a candidate which induces such infra-red (IR) effects. 
It is because the gravitational field is massless and contains minimally coupled modes \cite{Tsamis1992}. 

In order to investigate interacting field theories in dS space, 
we need to employ the Schwinger-Keldysh perturbation theory \cite{Schwinger1961,Keldysh1964}. 
The IR effects at each order manifest as the polynomials in the logarithm of the scale factor of the Universe: $\log a(\tau)$ \cite{Weinberg2005}. 
In an interacting scalar field theory with polynomial interactions such as $\lambda \varphi^4$ theory \cite{Woodard2002}, 
these IR logarithms in the propagators give rise to powers of IR logarithms in the amplitudes. 
Let us consider the expectation value of the energy-momentum tensor for example. 
The leading power of IR logarithms is given by the number of the propagators of the diagram. 
They make an effective cosmological constant time dependent and thus break the dS symmetry. 
Therefore we need to sum these leading IR logarithms to understand the long term evolution as the effect of IR logarithms becomes large if we wait long enough. 

Remarkably a simple physical picture holds in the leading IR log approximation as follows. 
A scalar field is not completely frozen beyond the horizon scale as it is constantly jolted by the modes coming out of the horizon. 
A scalar field performs a random walk in the scalar field space (1 dimension) which is consistent with the linear growth of the propagator with respect to cosmic time $t$ at the initial stage.\footnote{We recall here that the fractal dimension of random walk is $2$.}
Eventually it reaches an equilibrium state in a potential well as dS symmetry is restored \cite{Starobinsky1994,Woodard2005}. 
We have investigated IR logarithms in the non-linear sigma models which contain massless minimally coupled scalar fields with derivative interactions. 
In the expectation value of the energy-momentum tensor, we have shown that the leading IR logarithms cancel to all orders \cite{Kitamoto2011}. 

It is important to understand the IR effects in quantum gravity as the propagator of gravitons contains IR logarithms in dS space. 
The case is very strong here as gravitons exist in our Universe which is of dS type. 
In this paper we investigate physical effects of soft gravitons on microscopic physics in dS space. 
We focus on the dynamics of matter fields with sub-horizon momentum scale as it is directly observable. 
We find that the super-horizon gravitons influence the matter field dynamics inside the cosmological horizon. 
Our predictions are thus verifiable by direct observations in principle. 

We adopt a massless and conformally coupled scalar field and a massless Dirac field as matter field components. 
At the tree level, these fields respect the Lorentz invariance after the conformal transformation. 
In investigating soft graviton effects to local matter field dynamics, it is a non-trivial question whether they preserve the Lorentz invariance. 
We investigate the kinetic terms of matter fields, specifically the relative weight between time derivative term and  spatial derivative term to verify the Lorentz invariance.  
Furthermore in interacting field theories, we investigate how soft gravitons influence their coupling constants. 
As specific examples, we adopt $\lambda\phi^4$ and Yukawa theory with dimensionless couplings. 
The investigation in this paper is up to the one-loop level. 

E. O. Kahya and R. P. Woodard investigate soft graviton effects in a free massless and minimally coupled scalar field theory \cite{Kahya2007(1),Kahya2007(2)}. 
S. B. Giddings and M. S. Sloth investigate the scalar two point function in the quenched approximation \cite{Giddings2010}. 
S. P. Miao and R. P. Woodard compute the free Dirac field equation corrected by soft gravitons \cite{Miao2005(1)} 
and derive its solution at the super-horizon scale \cite{Miao2005(2),Miao2005(3)}.

In this paper we adopt the following three approximations: 
(i) We work in $D=4$ dimension as we focus on IR effects at the one-loop level. 
Known quantum effects such as conformal anomaly do not spoil dS symmetry.
It is because they are short distance effects and hence time independent for a fixed physical momentum scale.
(ii) We retain the dS symmetry breaking part of the amplitudes. 
As we focus on conformally coupled matter dynamics in this paper, we only retain the $\log a(\tau)$ part of the graviton propagators. 
(iii) Furthermore we neglect differentiated graviton propagators as they contain no $\log a(\tau)$ part. 
The approximation method has been introduced in Yukawa theory and scalar QED \cite{Duffy2005,Woodard2006,Prokopec2007}. 
However there is a caveat in trusting these three approximations as follows. 

Miao and Woodard have found that the on-shell quantum equation is sensitive to ultra-violet (UV) regularization 
and the approximations to retain only dS symmetry breaking part of the amplitudes fail \cite{Miao2005(2),Miao2005(3)}. 
It is an intriguing result which merits further investigations. 
We thought that the dS symmetry breaking effects are of IR origin. If so, why could they depend on UV regularizations? 
The puzzle may be resolved by distinguishing the dS symmetry breaking from the IR singularities in the on-shell limit. 
We investigate off-shell quantum equation to regulate such singularities which is a standard strategy in field theory. 
It is because the virtuality replaces the initial conformal time as the IR cut-off when we integrate over conformal time of the interaction vertices. 
The result is time independent as we fix the physical virtuality to be constant such as the confining scale in QCD for example. 
We can thus confine the source of the IR logarithms to the explicit dS symmetry breaking part of the soft graviton propagators. 
We need more work to extract physical quantities from the off-shell quantum equation.
In this respect, the virtuality may be replaced by energy resolution if we consider possible cancellation of IR singularities among energetically degenerate states.

The organization of this paper is as follows. 
In Section 2, we quantize the gravitational field on the dS background. We identify the graviton modes which exhibit the IR logarithm. 
In the subsequent sections, we evaluate the quantum equation of motion with respect to matter fields which are dressed by soft gravitons at the one-loop level. 
In Section 3, we adopt free field theories and investigate whether the Lorentz invariance is preserved. 
In Section 4, we evaluate the effective coupling constants in $\phi^4$ and Yukawa theory. 
We find that the IR effects from gravitons preserve the Lorentz invariance. 
The effective coupling constants are found to decrease with cosmic expansion with definite scaling exponents. 
In Section 5, we vary the gauge parameter of the graviton propagator to investigate the gauge dependences of the results obtained in Section 3 and 4. 
We show that the relative scaling exponents of the couplings are gauge invariant and observable. We conclude with discussions in Section 6.

\section{Gravitational field in de Sitter space}
\setcounter{equation}{0}

In this section, we compute the gravitation propagator in de Sitter (dS) space.  
In the Poincar\'{e} coordinate, the metric in dS space is
\begin{align}
ds^2=-dt^2+a^2(t)d{\bf x}^2,\hspace{1em}a(t)=e^{Ht}, 
\end{align}
where the dimension of dS space is taken as $D=4$ and $H$ is the Hubble constant. 
In the conformally flat coordinate,
\begin{align}
(g_{\mu\nu})_\text{dS}=a^2(\tau)\eta_{\mu\nu},\hspace{1em}a(\tau)=-\frac{1}{H\tau}. 
\end{align}
Here the conformal time $\tau\ (-\infty <\tau < 0)$ is related to the cosmic time $t$ as $\tau\equiv-\frac{1}{H}e^{-Ht}$. 
We assume that dS space begins at an initial time $t_i$ with a finite spacial extension.
After a sufficient exponential expansion, the dS space is well described locally by the above metric
irrespective of the spacial topology.
The metric is invariant under the scaling transformation 
\begin{align}
\tau\to C\tau,\hspace{1em}x^i\to Cx^i. 
\label{scaling}\end{align}
It is a part of the $SO(1,4)$ dS symmetry.

In dealing with the quantum fluctuation whose background is dS space, 
we adopt the following parametrization: 
\begin{align}
g_{\mu\nu}=\Omega^2(x)\tilde{g}_{\mu\nu},\ \Omega(x)=a(\tau)e^{\kappa w(x)}, 
\label{para1}\end{align}
\begin{align}
\det \tilde{g}_{\mu\nu}=-1,\ \tilde{g}_{\mu\nu}=(e^{\kappa h(x)})_{\mu\nu}, 
\label{para2}\end{align}
where $\kappa$ is defined by the Newton's constant $G$ as $\kappa^2=16\pi G$. 
To satisfy (\ref{para2}), $h_{\mu\nu}$ is traceless
\begin{align}
\eta^{\mu\nu}h_{\mu\nu}=0. 
\label{para3}\end{align}
By using this parametrization, the components of the Einstein action are written as follows. 
We keep a parameter  $D$ to specify the dimension for generality: 
\begin{align}
\sqrt{-g}=\Omega^D, 
\label{com1}\end{align}
\begin{align}
R=\Omega^{-2}\tilde{R}-2(D-1)\Omega^{-3}\tilde{g}^{\mu\nu}\nabla_\mu\partial_\nu\Omega-(D-1)(D-4)\Omega^{-4}\tilde{g}^{\mu\nu}\partial_\mu\Omega\partial_\nu\Omega, 
\label{com2}\end{align}
where $\tilde{R}$ is the Ricci scalar constructed from $\tilde{g}_{\mu\nu}$
\begin{align}
\tilde{R}=-\partial_\mu\partial_\nu\tilde{g}^{\mu\nu}
-\frac{1}{4}\tilde{g}^{\mu\nu}\tilde{g}^{\rho\sigma}\tilde{g}^{\alpha\beta}\partial_\mu\tilde{g}_{\rho\alpha}\partial_\nu\tilde{g}_{\sigma\beta}
+\frac{1}{2}\tilde{g}^{\mu\nu}\tilde{g}^{\rho\sigma}\tilde{g}^{\alpha\beta}\partial_\mu\tilde{g}_{\sigma\alpha}\partial_\rho\tilde{g}_{\nu\beta}. 
\label{com3}\end{align}
From (\ref{com1}) and (\ref{com2}), the Lagrangian of gravity is
\begin{align}
\mathcal{L}_\text{gravity}&=\frac{1}{\kappa^2}\sqrt{-g}\big[R-(D-1)(D-2)H^2\big]\label{gravity}\\
&=\frac{1}{\kappa^2}\big[\Omega^{D-2}\tilde{R}
+(D-1)(D-2)\Omega^{D-4}\tilde{g}^{\mu\nu}\partial_\mu\Omega\partial_\nu\Omega-(D-1)(D-2)H^2\Omega^{D}\big]. \notag
\end{align}
Note that the Lagrangian density is defined including $\sqrt{-g}$ in this paper. 

In order to fix the gauge degrees from general coordinate invariance
\begin{align}
x'_\mu&=x_\mu+\varepsilon_\mu, \\
g'_{\mu\nu}&=g_{\mu\nu}-g_{\mu\rho}\partial_\nu\varepsilon^\rho-g_{\nu\rho}\partial_\mu\varepsilon^\rho-\partial_\rho g_{\mu\nu}\varepsilon^\rho, \notag
\end{align}
we introduce the gauge fixing term \cite{Tsamis1992} 
\begin{align}
\mathcal{L}_\text{GF}&=-\frac{1}{2}a^{D-2}\eta^{\mu\nu}F_\mu F_\nu, \label{GF}\\
F_\mu&=\partial_\rho h_\mu^{\ \rho}-(D-2)\partial_\mu w+(D-2)h_\mu^{\ \rho}\partial_\rho\log a+2(D-2)w\partial_\mu\log a. \notag
\end{align}
The corresponding ghost term at the quadratic level is
\begin{align}
\mathcal{L}_\text{ghost}=&-a^{D-2}\partial_\sigma\bar{b}^\mu \eta^{\sigma\nu}
\big\{{\eta}_{\mu\rho}\partial_\nu+\eta_{\nu\rho}\partial_\mu+2\eta_{\mu\nu}\partial_\rho(\log a)\big\}b^\rho\label{ghost}\\
&+\partial_\mu(a^{D-2}\bar{b}^\mu)\eta^{\rho\sigma}
\big\{\eta_{\rho\nu}\partial_\sigma+\eta_{\rho\sigma}\partial_\nu(\log a)\big\}b^\nu, \notag
\end{align}
where $b^\mu$ is a ghost field and $\bar{b}^\mu$ is an anti-ghost field. 
From (\ref{com3}), (\ref{gravity}), (\ref{GF}) and (\ref{ghost}), the quadratic part of the total Lagrangian
density is
\begin{align}
\mathcal{L}_\text{quadratic}=&\ a^{D-2}\big\{\frac{1}{2}D(D-2)\eta^{\mu\nu}\partial_\mu w\partial_\nu w-\frac{D}{4(D-1)}\eta^{\mu\nu}\partial_\mu h^{00}\partial_\nu h^{00}
-\frac{1}{4}\eta^{\mu\nu}\partial_\mu\tilde{h}^i_{\ j}\partial_\nu\tilde{h}^j_{\ i}\notag\\
&\hspace{2.5em}+\frac{1}{2}\eta^{\mu\nu}\partial_\mu h^{0i}\partial_\nu h^{0i}
+\eta^{\mu\nu}\partial_\mu\bar{b}^0\partial_\nu b^0-\eta^{\mu\nu}\partial_\mu\bar{b}^i\partial_\nu b^i\big\}\notag\\
&+a^DH^2\big\{-2(D-2)w^2+2(D-2)w h^{00}-\frac{1}{2}(D-2)h^{00}h^{00}\label{quadratic}\\
&\hspace{4em}+\frac{1}{2}(D-2)h^{0i}h^{0i}+(D-2)\bar{b}^0b^0\big\}. \notag
\end{align}
Here we have decomposed $h^i_{\ j},\ i,j=1, \cdots, D-1$ into a trace part and a traceless part
\begin{align}
h^i_{\ j}=\tilde{h}^i_{\ j}+\frac{1}{D-1}h^k_{\ k}\delta^i_{\ j}=\tilde{h}^i_{\ j}+\frac{1}{D-1}h^{00}\delta^i_{\ j}. 
\end{align} 
(\ref{quadratic}) is diagonalized as
\begin{align}
\mathcal{L}_\text{quadratic}=a^D&\big[\ \frac{1}{2}a^{-2}\eta^{\mu\nu}\partial_\mu X\partial_\nu X-\frac{1}{4}a^{-2}\eta^{\mu\nu}\partial_\mu\tilde{h}^i_{\ j}\partial_\nu\tilde{h}^j_{\ i}
-a^{-2}\eta^{\mu\nu}\partial_\mu\bar{b}^i\partial_\nu b^i\label{quadratic1}\\
&+\frac{1}{2}a^{-2}\eta^{\mu\nu}\partial_\mu h^{0i}\partial_\nu h^{0i}+\frac{1}{2}(D-2)H^2h^{0i}h^{0i}\notag\\
&-\frac{1}{2}a^{-2}\eta^{\mu\nu}\partial_\mu Y\partial_\nu Y-(D-3)H^2Y^2\notag\\
&+a^{-2}\eta^{\mu\nu}\partial_\mu\bar{b}^0\partial_\nu b^0+(D-2)H^2\bar{b}^0b^0\big], \notag
\end{align}
where $X,\ Y$ are
\begin{align}
X=(D-2)\sqrt{\frac{D-1}{D-3}}w-\frac{1}{\sqrt{(D-1)(D-3)}}h^{00},\hspace{1em}Y=\sqrt{\frac{D-2}{2(D-3)}}(h^{00}-2w). 
\label{diagonalize}\end{align}
(\ref{quadratic1}) contains three types of fields. 
One is a massless and minimally coupled field: $X,\tilde{h}^i_{\ j}$, $b^i$, $\bar{b}^i$. 
The others are two types of massless and non-minimally coupled fields: $h^{0i}$, $b^0$, $\bar{b}^0$, and $Y$. 

We restrict to the $D=4$ case in the subsequent discussion. 
In this case, the non-minimally coupled fields correspond with a massless conformally coupled field. 
We list gravitation propagators as below
\begin{align}
\langle X(x)X(x')\rangle&=-\langle\varphi(x)\varphi(x')\rangle, \label{minimally}\\
\langle\tilde{h}^i_{\ j}(x)\tilde{h}^k_{\ l}(x')\rangle&=(\delta^{ik}\delta_{jl}+\delta^i_{\ l}\delta_j^{\ k}-\frac{2}{3}\delta^i_{\ j}\delta^k_{\ l})\langle\varphi(x)\varphi(x')\rangle, \notag\\
\langle b^i(x)\bar{b}^j(x')\rangle&=\delta^{ij}\langle\varphi(x)\varphi(x')\rangle, \notag
\end{align}
\begin{align}
\langle h^{0i}(x)h^{0j}(x')\rangle&=-\delta^{ij}\langle\phi(x)\phi(x')\rangle, \label{conformally}\\
\langle Y(x)Y(x')\rangle&=\langle\phi(x)\phi(x')\rangle, \notag\\
\langle b^0(x)\bar{b}^0(x')\rangle&=-\langle\phi(x)\phi(x')\rangle, \notag
\end{align}
Here $\varphi$ denotes a massless and minimally coupled scalar field and $\phi$ denotes a massless conformally coupled scalar field
\begin{align}
\langle\varphi(x)\varphi(x')\rangle&=\frac{H^2}{4\pi^2}\big\{\frac{1}{y}-\frac{1}{2}\log y+\frac{1}{2}\log a(\tau)a(\tau')+1-\gamma\big\}, 
\label{minimally0}\end{align} 
\begin{align}
\langle\phi(x)\phi(x')\rangle&=\frac{H^2}{4\pi^2}\frac{1}{y}, 
\label{conformally0}\end{align}
where $\gamma$ is Euler's constant and $y$ is the dS invariant distance
\begin{align}
y=\frac{-(\tau-\tau')^2+({\bf x}-{\bf x}')^2}{\tau\tau'}. 
\end{align}
It should be noted that the propagator for a massless and minimally coupled scalar field has 
the dS symmetry breaking logarithmic term: $\log a(\tau)a(\tau')$. 
To explain what causes the dS symmetry breaking, 
we recall the wave function for a massless and minimally coupled field
\begin{align}
\phi_{\bf p}(x)=\frac{H\tau}{\sqrt{2p}}(1-i\frac{1}{p\tau})e^{-ip\tau+i{\bf p}\cdot{\bf x}}. 
\end{align}
Inside the cosmological horizon, the physical momentum: $P\equiv p/a(\tau)\gg H \Leftrightarrow p|\tau|\gg 1$, 
this wave function approaches that in Minkowski space up to a cosmic scale factor
\begin{align}
\phi_{\bf p}(x)\sim\frac{H\tau}{\sqrt{2p}}e^{-ip\tau+i{\bf p}\cdot{\bf x}}. 
\end{align}
On the other hand, the behavior outside the cosmological horizon $P\ll H$ is
\begin{align}
\phi_{\bf p}(x)\sim\frac{H}{\sqrt{2p^3}}e^{i{\bf p}\cdot{\bf x}}. 
\end{align}
The IR behavior indicates that the corresponding propagator has a scale invariant spectrum. 
As a direct consequence of it, the propagator has a logarithmic divergence from the IR contributions
in the infinite volume limit. 

To regularize the IR divergence, we introduce an IR cut-off $\varepsilon_0$ which fixes the minimum value of the comoving momentum 
\begin{align}
\int^H_{\varepsilon_0a^{-1}(\tau)} dP. 
\end{align}
With this prescription, the degrees of freedom (d.o.f.) outside the cosmological horizon increase with cosmic evolution. 
Due to the increase, the propagator acquires a growing time dependence which spoils the dS symmetry. 
In tribute to its origin, we call the dS symmetry breaking term the IR logarithm. 
Physically speaking, $1/\varepsilon_0$ is recognized as an initial size of the Universe when the exponential expanding starts. 
For simplicity, we set $\varepsilon_0=H$ in (\ref{minimally0}). 

As there is explicit  time dependence in the propagator, 
physical quantities can acquire time dependence through the quantum loop corrections. 
We call them the quantum IR effects in dS space. 
By the power counting of the IR logarithms in quantum gravity, 
the leading IR effects at $n$-loop level are estimated as $(\kappa^2H^2\log a(\tau))^n$. 
The estimation indicates that even if $\kappa^2H^2\ll 1$, the quantum effects can eventually grow up to the tree level magnitude . 
This is the reason why we focus on the quantum IR effects in dS space. 
Before concluding this section, we introduce an approximation. 
Focusing on the IR effects, 
we can neglect conformally coupled modes of gravity since they do not induce the IR logarithm. 
By applying this approximation, we can identify the following two modes as
\begin{align}
h^{00}\simeq 2w\simeq\frac{\sqrt{3}}{2}X. 
\label{diagonalize1}\end{align}

\section{Quantum equation of motion}
\setcounter{equation}{0}

In the preceding section, we have reviewed that the gravitational field contains massless and minimally coupled modes 
and the corresponding propagator is time dependent due to the increase of d.o.f. outside the cosmological horizon. 
In this section, we investigate how the quantum IR effects from gravitons influence the local dynamics of the matter fields. 
More specifically, we evaluate the quantum equation of motion for matter fields including the quantum IR effects from gravitons. 

To begin with, we review how to derive the quantum equation of motion on a time dependent background \cite{Hu1997}. 
Let us represent the vacuum at $t\to-\infty$ as $|in\rangle$, and $t\to\infty$ as $|out\rangle$. 
In the Feynman-Dyson formalism on a flat background, 
it is presumed that $|out\rangle$ is equal to $|in\rangle$ up to a phase factor. 
On the other hand, we cannot prefix $|out\rangle$ in dS space. 
The correct strategy is to evaluate vacuum expectation values (vev) with respect to $|in\rangle$: 
\begin{align}
\langle \mathcal{O}_H(x)\rangle
=\langle in| T_C[U(-\infty,\infty)U(\infty,-\infty)\mathcal{O}_I(x)]|in\rangle. 
\end{align}
where $\mathcal{O}_H$ and $\mathcal{O}_I$ denote the operators in the Heisenberg and the interaction pictures respectively. 
$U(t_1,t_2)$ is the time translation operator in the interaction picture 
\begin{align}
U(t_1,t_2)=\exp\big\{i\int^{t_1}_{t_2}d^4x\ \delta\mathcal{L}_I(x)\big\}. 
\end{align}
$\delta \mathcal{L}$ denotes the interaction term of the Lagrangian.  
It is crucial that the operator ordering $T_C$ specified by the following path is adopted here
\begin{align}
&\parbox{\SKpathl}{\usebox{\SKpath}}, \\
&\hspace{3em}\int_C dt = \int^\infty_{-\infty} dt_+ - \int^\infty_{-\infty} dt_-. \notag
\end{align}
We call it the Schwinger-Keldysh formalism. 
Since there are two time indices $+,-$ in this formalism, the propagator has four components
\begin{align}
\begin{pmatrix} \langle\varphi_+(x)\varphi_+(x')\rangle & \langle\varphi_+(x)\varphi_-(x')\rangle \\
\langle\varphi_-(x)\varphi_+(x')\rangle & \langle\varphi_-(x)\varphi_-(x')\rangle \end{pmatrix}
=\begin{pmatrix} \langle T\varphi(x)\varphi(x')\rangle & \langle\varphi(x')\varphi(x)\rangle \\
\langle\varphi(x)\varphi(x')\rangle & \langle\tilde{T}\varphi(x)\varphi(x')\rangle \end{pmatrix}, 
\label{4propagators}\end{align}
where $\tilde{T}$ denotes the anti-time ordering. 

We introduce the external source $J_+,J_-$ for each path and evaluate
\begin{align}
Z[J_+,J_-]=
\langle in| T_C[U(-\infty,\infty)U(\infty,-\infty)
\exp\big\{i\int d^4x\ (J_+\varphi_+ -J_-\varphi_-)\big\}
]|in\rangle. 
\end{align}
The generating functional for the connected Green's functions is
\begin{align}
iW[J_+,J_-]=\log Z[J_+,J_-]. 
\label{connected}\end{align}
We define the classical field as
\begin{align}
\hat{\varphi}_A(x)=c_{AB}\frac{\delta W[J_+,J_-]}{\delta J_B(x)},\hspace{1em}A,B=+,-, 
\label{varphi}\end{align}
\begin{align}
c_{AB}=\begin{pmatrix} 1 & 0 \\ 0 & -1\end{pmatrix}. 
\end{align}
By taking the limit $J_+=J_-=J$ in (\ref{varphi}), 
we obtain the vev of $\varphi$ where the action contains the additional $J\varphi$ term 
\begin{align}
\langle \varphi(x)\rangle|_{J\varphi}=\hat{\varphi}_+(x)|_{J_+=J_-=J}=\hat{\varphi}_-(x)|_{J_+=J_-=J}. 
\label{coincident1}\end{align}
Finally, we turn off the source term $J=0$ 
\begin{align}
\langle\varphi(x)\rangle=\frac{\delta W[J_+,J_-]}{\delta J_+(x)}\Big|_{J_+=J_-=0}
=-\frac{\delta W[J_+,J_-]}{\delta J_-(x)}\Big|_{J_+=J_-=0}. 
\label{vev}\end{align}

The effective action is obtained after the Legendre transformation
\begin{align}
\Gamma[\hat{\varphi}_+,\hat{\varphi}_-]=W[J_+,J_-]-\int d^4x\ (J_+\hat{\varphi}_+ -J_-\hat{\varphi}_-), 
\end{align}
where $J_{+,-}$ are given by $\hat{\varphi}_{+,-}$ as follows
\begin{align}
J_A(x)=-c_{AB}\frac{\delta \Gamma[\hat{\varphi}_+,\hat{\varphi}_-]}{\delta \hat{\varphi}_B(x)}. 
\label{J}\end{align} 
From (\ref{coincident1}) and (\ref{J}), we obtain in the limit $\hat{\varphi}_+=\hat{\varphi}_-=\hat{\varphi}$
\begin{align}
J(x)=-\frac{\delta \Gamma[\hat{\varphi}_+,\hat{\varphi}_-]}{\delta \hat{\varphi}_+(x)}\Big|_{\hat{\varphi}_+=\hat{\varphi}_-=\hat{\varphi}}
=\frac{\delta \Gamma[\hat{\varphi}_+,\hat{\varphi}_-]}{\delta \hat{\varphi}_-(x)}\Big|_{\hat{\varphi}_+=\hat{\varphi}_-=\hat{\varphi}}. 
\label{coincident2}\end{align}
In the absence of the external source, the exact equation of motion is obtained including quantum effects
\begin{align}
\frac{\delta \Gamma[\hat{\varphi}_+,\hat{\varphi}_-]}{\delta \hat{\varphi}_+(x)}\Big|_{\hat{\varphi}_+=\hat{\varphi}_-=\hat{\varphi}}
=-\frac{\delta \Gamma[\hat{\varphi}_+,\hat{\varphi}_-]}{\delta \hat{\varphi}_-(x)}\Big|_{\hat{\varphi}_+=\hat{\varphi}_-=\hat{\varphi}}=0. 
\label{EEoM}\end{align}
We call these identities the quantum equation of motion in this paper.
In the following subsections, we evaluate these identities in concrete models to understand quantum IR effects from gravitons to matter field dynamics at the one-loop level. 

\subsection{Free field theories}

Let us investigate the effects of soft gravitons on the local dynamics of a free scalar and Dirac field.
First we investigate a massless conformally coupled scalar field. 
The corresponding action is 
\begin{align}
S=\int \sqrt{-g}d^4x\big[-\frac{1}{2}g^{\mu\nu}\partial_\mu\phi\partial_\nu\phi-\frac{1}{12}R\phi^2\big]. 
\end{align}
For convenience, we redefine the matter field: 
\begin{align}
\Omega\phi\to\phi, 
\label{frs}\end{align}
\begin{align}
S=\int d^4x\big[-\frac{1}{2}\tilde{g}^{\mu\nu}\partial_\mu\phi\partial_\nu\phi-\frac{1}{12}\tilde{R}\phi^2\big]. 
\label{scalar}\end{align}
To obtain the quantum equation of the matter field, 
we decompose it into the classical field and the quantum fluctuation
\begin{align}
\phi\to\hat{\phi}+\phi. 
\label{sdivide}\end{align} 
By differentiating (\ref{scalar}) with respect to $\hat{\phi}_+$ as in (\ref{EEoM}),
we obtain the quantum equation of motion. 
The quantum equation of motion up to the one-loop level is expressed as 
\begin{align}
&\partial^2\hat{\phi}(x)+\frac{1}{2}\kappa^2\partial_\mu(\langle (h^{\mu\rho})_+(x)(h_\rho^{\ \nu})_+(x)\rangle\partial_\nu\hat{\phi}(x))\label{sEoM1}\\
&-i\kappa^2\partial_\mu\int d^4x'\ c_{AB}\langle (h^{\mu\nu})_+(x)(h^{\rho\sigma})_A(x')\rangle\langle\partial_\nu\phi_+(x)\partial_\rho'\phi_B(x')\rangle\partial_\sigma'\hat{\phi}(x')
\simeq 0, \notag
\end{align}
where $\partial^2=\eta^{\mu\nu}\partial_\mu\partial_\nu$ and 
we have taken the limit $\hat{\phi}_+=\hat{\phi}_-=\hat{\phi}$. 

We have neglected the Ricci scalar in (\ref{sEoM1}) and do not consider the contribution from differentiated gravitational fields also in the subsequent calculations. 
The differentiated gravitons lead to the following terms in the quantum equation
\begin{align}
(\partial\log a(\tau))\partial\hat{\phi}\sim Ha(\tau)\partial\hat{\phi}(x),\hspace{1em}
(\partial\partial\log a(\tau))\hat{\phi}\sim H^2a^2(\tau)\hat{\phi}(x). 
\label{DG}\end{align}
In this paper, we investigate the effects of the IR logarithms due to soft gravitons on the microscopic matter dynamics. 
Specifically we set the external momentum to be of the sub-horizon scale: 
\begin{align}
P\gg H\ \Leftrightarrow\ \partial\hat{\phi}(x)\gg Ha(\tau)\hat{\phi}(x), 
\label{sub}\end{align}
and focus on the super-horizon fluctuation of the internal momentum: 
\begin{align}
\langle h^{\mu\nu}(x)h^{\rho\sigma}(x)\rangle \to H^2\int^\frac{-1}{\tau}_H\frac{dq}{q}=H^2\log a(\tau). 
\label{super}\end{align}
Note that $\partial$ operating the classical field is identified as the external comoving momentum scale $p$ 
and the conformally coupled scalar field does not induce the IR logarithm. 

The dynamics at the sub-horizon scale is directly observable. 
That is why we adopt the setting of the external momentum (\ref{sub}). 
Of course we also assume the external momentum to be much smaller than the Planck scale such that quantum gravity effect is small. 
The remarkable point is that the IR logarithmic correction focused in (\ref{super}) may become large 
at late times to be of the same magnitude of the tree order. 
The other terms neglected in (\ref{super}) lead to UV divergences in general. 
We can renormalize them by introducing all possible counter terms. 
Although there are infinite freedoms to choose finite parts of the counter terms, 
the UV corrections are constant in time contrary to the IR corrections and so suppressed by $\kappa^2H^2$ or $\kappa^2P^2$. 
If we impose only the condition (\ref{sub}), we need to consider higher derivative terms. 
However the IR logarithms do not associate with them. 
The requirement of (\ref{sub}) and (\ref{super}) indicates that we evaluate the coefficient of the following term
\begin{align}
\kappa^2H^2\log a(\tau)\partial\partial\hat{\phi}(x). 
\end{align}
That is why we do not consider the contribution from differentiated gravitational fields in this paper. 
Of course, if we are interested in the dynamics at the super-horizon scale, 
it is not correct to neglect differentiated gravitational fields. 
As an example, please see Eq. (90)-(91) in \cite{Miao2005(2)}. 
We also refer to the fact that as far as we consider the dynamics at the sub-horizon scale, 
there is no loss of generality to assume that the scalar field is conformally coupled to gravity
\footnote{In the minimally coupled case, we need to include the contribution from soft scalar and hard graviton intermediate states.}.

From (\ref{minimally}) and (\ref{diagonalize1}), 
we have only to focus on the following propagators to extract massless and minimally coupled modes from gravity
\begin{align}
\langle h^{00}(x)h^{00}(x')\rangle&\simeq-\frac{3}{4}\langle\varphi(x)\varphi(x')\rangle, \label{gravityp}\\
\langle h^{00}(x)h^i_{\ j}(x')\rangle&\simeq-\frac{1}{4}\delta^i_{\ j}\langle\varphi(x)\varphi(x')\rangle, \notag\\
\langle h^i_{\ j}(x)h^k_{\ l}(x')\rangle&\simeq(\delta^{ik}\delta_{jl}+\delta^i_{\ l}\delta_j^{\ k}-\frac{3}{4}\delta^i_{\ j}\delta^k_{\ l})\langle\varphi(x)\varphi(x')\rangle. \notag
\end{align}
By adopting an UV regularization, 
the propagator at coincident point is estimated as follows
\begin{align}
\langle\varphi(x)\varphi(x)\rangle=\text{(UV divergent const)}+\frac{H^2}{4\pi^2}\log a(\tau). 
\end{align}
Along the above discussion, we focus on the time dependent dS symmetry breaking part 
\begin{align}
\langle\varphi(x)\varphi(x)\rangle\simeq\frac{H^2}{4\pi^2}\log a(\tau). 
\label{minimallyc}\end{align}
On the other hand, the matter field contains only a conformally coupled mode and so we can use the
exact propagator
\begin{align}
\langle \phi(x)\phi(x')\rangle=\frac{1}{4\pi^2}\frac{1}{\Delta x^2}, 
\label{conformally1}\end{align}
where $\Delta x^\mu\equiv x^\mu-x'^\mu,\ \Delta x^2\equiv\eta_{\mu\nu}\Delta x^\mu\Delta x^\nu$. 
It should be noted that we have redefined the matter field.  
In (\ref{minimally0}) and (\ref{conformally1}), 
the Schwinger-Keldysh indices are assigned as follows
\begin{align}
y_{AB}&=H^2a(\tau)a(\tau')\Delta x_{AB}^2,\hspace{1em}A,B=\pm, 
\end{align}
\begin{align}
\Delta x_{++}^2&=-(|\tau-\tau'|-ie)^2+({\bf x}-{\bf x}')^2, \label{pm}\\
\Delta x_{+-}^2&=-(\tau-\tau'+ie)^2+({\bf x}-{\bf x}')^2, \notag\\
\Delta x_{-+}^2&=-(\tau-\tau'-ie)^2+({\bf x}-{\bf x}')^2, \notag\\
\Delta x_{--}^2&=-(|\tau-\tau'|+ie)^2+({\bf x}-{\bf x}')^2, \notag
\end{align} 
where $e$ is a positive infinitesimal quantity. 

From (\ref{sub}), (\ref{gravityp}) and (\ref{minimallyc}), 
the second term in (\ref{sEoM1}) is evaluated as
\begin{align}
\frac{1}{2}\kappa^2\partial_\mu(\langle (h^{\mu\rho})_+(x)(h_\rho^{\ \nu})_+(x)\rangle\partial_\nu\hat{\phi}(x))
\simeq\frac{\kappa^2H^2}{4\pi^2}\log a(\tau)\big\{\frac{3}{8}\partial_0^2+\frac{13}{8}\partial_i^2\big\}\hat{\phi}(x). 
\label{sEoM2}\end{align}
To evaluate the third term in (\ref{sEoM1}), we need to perform the following integration
\begin{align}
&-i\kappa^2\partial_\mu\int d^4x'\ c_{AB}\langle \varphi_+(x)\varphi_A(x')\rangle\langle\partial_\nu\phi_+(x)\partial_\rho'\phi_B(x')\rangle\partial_\sigma'\hat{\phi}(x')\label{sint1}\\
=&\ i\kappa^2\partial_\mu\int d^4x'\ c_{AB}\partial_\sigma'\big\{\langle \varphi_+(x)\varphi_A(x')\rangle\langle\partial_\nu\phi_+(x)\partial_\rho'\phi_B(x')\rangle\big\}\hat{\phi}(x'). \notag
\end{align} 
We should note that the dS invariant part of $\langle \varphi_+(x)\varphi_A(x')\rangle$ 
does not contribute to the coefficient of $\log a(\tau)$. 
It is because under the scaling transformation (\ref{scaling}), 
the corresponding integral scales in agreement with its dimension: 
\begin{align}
&i\kappa^2\partial_\mu\int d^4x'\ c_{AB}\partial_\sigma'\big\{F(y_{+A})\langle\partial_\nu\phi_+(x)\partial_\rho'\phi_B(x')\rangle\big\}\hat{\phi}(x') \label{general}\\
\to &\ C^{-2}\times i\kappa^2\partial_\mu\int d^4x'\ c_{AB}\partial_\sigma'\big\{F(y_{+A})
\langle\partial_\nu\phi_+(x)\partial_\rho'\phi_B(x')\rangle\big\}\hat{\phi}(Cx'), \notag
\end{align} 
where $F$ is a certain function.  
In order to evaluate the coefficient of $\log a(\tau)$, we only need to retain the dS breaking term 
of $\langle \varphi_+(x)\varphi_A(x')\rangle$:
\begin{align}
\langle\varphi(x)\varphi(x')\rangle\simeq\frac{H^2}{8\pi^2}\log a(\tau)a(\tau'). 
\label{minimallya}\end{align}
This breaking term does not contain the Schwinger-Keldysh index. 
Up to $\mathcal{O}(\log a(\tau))$, the following identity works
\begin{align}
&i\kappa^2\partial_\mu\int d^4x'\ \partial_\sigma'\big\{\frac{H^2}{8\pi^2}\log a(\tau)a(\tau') \big[\langle\partial_\nu\phi_+(x)\partial_\rho'\phi_+(x')\rangle
-\langle\partial_\nu\phi_+(x)\partial_\rho'\phi_-(x')\rangle\big]\big\}\hat{\phi}(x')\label{out1}\\
\simeq&\ \frac{H^2}{4\pi^2}\log a(\tau)\times i\kappa^2\partial_\mu\partial_\nu\partial_\rho\partial_\sigma\int d^4x'\ \big[\langle\phi_+(x)\phi_+(x')\rangle-\langle\phi_+(x)\phi_-(x')\rangle\big]\hat{\phi}(x'). \notag
\end{align}
Here we have neglected the differentiated logarithms and used the translation symmetry of $\langle\phi(x)\phi(x')\rangle$. 
The above approximation method has been introduced in Yukawa theory and scalar QED \cite{Duffy2005,Woodard2006,Prokopec2007}. 

To evaluate the kinetic term, we need to  expand $\hat{\phi}$ up to the second order
\begin{align}
\hat{\phi}(x')\to\hat{\phi}(x)-\partial_\alpha\hat{\phi}(x)\Delta x^\alpha+\frac{1}{2}\partial_\alpha\partial_\beta\hat{\phi}(x)\Delta x^\alpha\Delta x^\beta. 
\label{sexpansion}\end{align}
From (\ref{conformally1}), (\ref{out1}) and (\ref{sexpansion}), (\ref{sint1}) is written as 
\begin{align}
i\frac{\kappa^2H^2}{16\pi^4}\log a(\tau)\partial_\mu\partial_\nu\partial_\rho\partial_\sigma
\int d^4x'&\Big\{\big[\frac{1}{\Delta x_{++}^2}-\frac{1}{\Delta x_{+-}^2}\big]\hat{\phi}(x)
-\big[\frac{\Delta x^\alpha}{\Delta x_{++}^2}-\frac{\Delta x^\alpha}{\Delta x_{+-}^2}\big]\partial_\alpha\hat{\phi}(x)\\
&+\frac{1}{2}\big[\frac{\Delta x^\alpha\Delta x^\beta}{\Delta x_{++}^2}-\frac{\Delta x^\alpha\Delta x^\beta}{\Delta x_{+-}^2}\big]
\partial_\alpha\partial_\beta\hat{\phi}(x)\Big\}. \notag
\end{align}

To investigate the kinetic term, we need to perform the following integrations. 
We list the results below 
\begin{align}
\partial_\alpha\partial_\beta\int d^4x'\big[\frac{1}{\Delta x_{++}^2}-\frac{1}{\Delta x_{+-}^2}\big]
=-4i\pi^2\delta_\alpha^{\ 0}\delta_\beta^{\ 0}, 
\label{C1}\end{align}
\begin{align}
\partial_\beta\partial_\gamma\partial_\delta\int d^4x'\big[\frac{\Delta x_\alpha}{\Delta x_{++}^2}-\frac{\Delta x_\alpha}{\Delta x_{+-}^2}\big]
= 8i\pi^2\delta_\alpha^{\ 0}\delta_\beta^{\ 0}\delta_\gamma^{\ 0}\delta_\delta^{\ 0}, 
\label{C2}\end{align}
\begin{align}
&\partial_\gamma\partial_\delta\partial_\varepsilon\partial_\eta\int d^4x'\big[\frac{\Delta x_\alpha\Delta x_\beta}{\Delta x_{++}^2}-\frac{\Delta x_\alpha\Delta x_\beta}{\Delta x_{+-}^2}\big]\label{C3}\\
=&-32i\pi^2\delta_\alpha^{\ 0}\delta_\beta^{\ 0}\delta_\gamma^{\ 0}\delta_\delta^{\ 0}\delta_\varepsilon^{\ 0}\delta_\eta^{\ 0}
-8i\pi^2\eta_{\alpha\beta}\delta_\gamma^{\ 0}\delta_\delta^{\ 0}\delta_\varepsilon^{\ 0}\delta_\eta^{\ 0}. \notag
\end{align}
We explain how to derive them in Appendix \ref{A:A}. 
In total, (\ref{sint1}) induces the following kinetic term with the IR logarithm
\begin{align}
\frac{\kappa^2H^2}{4\pi^2}\log a(\tau)\Big\{&
\delta_\mu^{\ 0}\delta_\nu^{\ 0}\partial_\rho\partial_\sigma+\delta_\mu^{\ 0}\delta_\rho^{\ 0}\partial_\nu\partial_\sigma+\delta_\mu^{\ 0}\delta_\sigma^{\ 0}\partial_\nu\partial_\rho
+\delta_\nu^{\ 0}\delta_\rho^{\ 0}\partial_\mu\partial_\sigma+\delta_\nu^{\ 0}\delta_\sigma^{\ 0}\partial_\mu\partial_\rho+\delta_\rho^{\ 0}\delta_\sigma^{\ 0}\partial_\mu\partial_\nu\notag\\
&-2(\delta_\mu^{\ 0}\delta_\nu^{\ 0}\delta_\rho^{\ 0}\partial_\sigma+\delta_\mu^{\ 0}\delta_\nu^{\ 0}\delta_\sigma^{\ 0}\partial_\rho
+\delta_\mu^{\ 0}\delta_\rho^{\ 0}\delta_\sigma^{\ 0}\partial_\nu+\delta_\nu^{\ 0}\delta_\rho^{\ 0}\delta_\sigma^{\ 0}\partial_\mu)\partial_0\notag\\
&+4\delta_\mu^{\ 0}\delta_\nu^{\ 0}\delta_\rho^{\ 0}\delta_\sigma^{\ 0}\partial_0^2+\delta_\mu^{\ 0}\delta_\nu^{\ 0}\delta_\rho^{\ 0}\delta_\sigma^{\ 0}\partial^2
\Big\}\hat{\phi}(x). \label{sint2}
\end{align}
In (\ref{C1})-(\ref{C3}), the integrands contain the step function $\theta(\tau-\tau')$ due to the causality. 
In fact, these integrals take finite values just from the derivative of the step function. So they are identified as the local terms.  
The above procedure implicitly indicates that only the local terms contribute to the dS symmetry breaking. 

Here we should emphasize that we investigate the off-shell effective field equation in this paper. 
In investigating the dS symmetry breaking, we need to distinguish the initial time dependence from the dependence of virtuality. 
The non-local terms respect the dS symmetry due to the existence of the nonzero virtuality. 
For more detail, please refer to Appendix \ref{A:B}. 

From (\ref{gravityp}) and (\ref{sint2}), the third term in (\ref{sEoM1}) leads to
\begin{align}
\frac{\kappa^2H^2}{4\pi^2}\log a(\tau)\big\{-\frac{3}{4}\partial_0^2-\frac{5}{4}\partial_i^2\big\}\hat{\phi}(x). 
\label{sEoM3}\end{align}
From (\ref{sEoM2}) and (\ref{sEoM3}), the quantum equation of motion of a scalar field including the one-loop correction from soft gravitons is
\begin{align}
\big\{1+\frac{3\kappa^2H^2}{32\pi^2}\log a(\tau)\big\}\partial^2\hat{\phi}(x). 
\label{soverall}\end{align}
Although each contribution (\ref{sEoM2}), (\ref{sEoM3}) breaks the Lorentz invariance, 
the total of them preserves it. 
The IR effect emerges just as an overall factor\footnote{In the minimally coupled case, the contribution from soft scalar and hard graviton intermediate states
enhances the quantum correction by a factor of 4/3.}. 
Since the derivative of $\log a(\tau)$ is negligible on the local dynamics at the sub-horizon scale, 
we can eliminate it by the following time dependent renormalization of a scalar field:
\begin{align}
\phi\to Z_\phi\phi,\hspace{1em}Z_\phi\simeq 1-\frac{3\kappa^2H^2}{64\pi^2}\log a(\tau). 
\label{sZ}\end{align}

We have checked that the same result is obtained in an exact calculation with the dimensional regularization. 
The IR logarithm originates from the dS symmetry breaking term in the graviton propagator in such a calculation.
For more detail, please refer to Appendix \ref{A:B}. 
We also remark that the IR logarithm can be absorbed into the wave function renormalization factor 
even if we include a mass term as a perturbation.

Next we perform a parallel investigation with a Dirac field. 
The corresponding action is 
\begin{align}
S=\int \sqrt{-g}d^4x\ i\bar{\psi}e^\mu_{\ a}\gamma^a \nabla_\mu\psi, 
\end{align}
where $e^\mu_{\ a}$ is a vierbein and $\gamma^a$ is the gamma matrix: 
\begin{align}
\gamma^a\gamma^b+\gamma^b\gamma^a=-2\eta^{ab}. 
\end{align}
The vierbein can be parametrized as
\begin{align}
e^\mu_{\ a}=\Omega^{-1} \tilde{e}^\mu_{\ a},\hspace{1em}\tilde{e}^\mu_{\ a}=(e^{-\frac{\kappa}{2}h})_{\ a}^{\mu}.
\end{align}
In a similar way to (\ref{scalar}), we redefine the matter field: 
\begin{align}
\Omega^\frac{3}{2}\psi\to\psi, 
\label{frD}\end{align} 
\begin{align}
S=\int d^4x\ i\bar{\psi}\tilde{e}^\mu_{\ a}\gamma^a \tilde{\nabla}_\mu\psi. 
\label{Dirac}\end{align}
By decomposing $\psi$ into the classical field and the quantum fluctuation
\begin{align}
\psi\to\hat{\psi}+\psi, 
\label{Ddivide}\end{align}
and differentiating (\ref{Dirac}) with respect to $\hat{\bar{\psi}}$, 
the quantum equation of motion up to the one-loop level is written as
\begin{align}
&i\eta^\mu_{\ a}\gamma^a\partial_\mu\hat{\psi}(x)+i\frac{\kappa^2}{8}\langle (h^\mu_{\ \rho})_+(x)(h^\rho_{\ a})_+(x)\rangle\gamma^a\partial_\mu\hat{\psi}(x)\label{DEoM1}\\
&-i\frac{\kappa^2}{4}\int d^4x'\ c_{AB}\langle (h^\mu_{\ a})_+(x)(h^\nu_{\ b})_A(x')\rangle\gamma^a\langle\partial_\mu\psi_+(x)\bar{\psi}_B(x')\rangle\gamma^b\partial_\nu'\hat{\psi}(x')
\simeq 0. \notag
\end{align} 
Here we have approximated $\tilde{\nabla}_\mu\simeq\partial_\mu$ 
since the spin connection consists of the the differentiated gravitational field. 

By substituting the identity $\langle\psi(x)\bar{\psi}(x')\rangle=i\eta^\rho_{\ c}\gamma^c\partial_\rho\langle\phi(x)\phi(x')\rangle$, (\ref{DEoM1}) is written as
\begin{align}
&i\eta^\mu_{\ a}\gamma^a\partial_\mu\hat{\psi}(x)+i\frac{\kappa^2}{8}\langle (h^\mu_{\ \rho})_+(x)(h^\rho_{\ a})_+(x)\rangle\gamma^a\partial_\mu\hat{\psi}(x)\label{DEoM2}\\
&+\frac{\kappa^2}{4}\eta^\rho_{\ c}\int d^4x'\ c_{AB}
\langle (h^\mu_{\ a})_+(x)(h^\nu_{\ b})_A(x')\rangle\gamma^a\gamma^c\langle\partial_\mu\partial_\rho\phi_+(x)\phi_B(x')\rangle\gamma^b\partial_\nu'\hat{\psi}(x')
\simeq 0. \notag
\end{align} 
From (\ref{gravityp}) and (\ref{minimallyc}), 
the second term in (\ref{DEoM2}) is evaluated as
\begin{align}
i\frac{\kappa^2}{8}\langle (h^\mu_{\ \rho})_+(x)(h^\rho_{\ a})_+(x)\rangle\gamma^a\partial_\mu\hat{\psi}(x)
\simeq \frac{\kappa^2H^2}{4\pi^2}\log a(\tau)\times i\big\{-\frac{3}{32}\gamma^0\partial_0+\frac{13}{32}\gamma^i\partial_i\big\}\hat{\psi}(x). 
\label{DEoM3}\end{align}
To evaluate the third term in (\ref{DEoM2}), we need to perform the following integration
\begin{align}
&\ \frac{\kappa^2}{4}\eta^\rho_{\ c}\int d^4x'\ c_{AB}
\langle \varphi_+(x)\varphi_A(x')\rangle\gamma^a\gamma^c\langle\partial_\mu\partial_\rho\phi_+(x)\phi_B(x')\rangle\gamma^b\partial_\nu'\hat{\psi}(x')\label{Dint1}\\
\simeq&\ \frac{\kappa^2}{4}\frac{H^2}{4\pi^2}\log a(\tau)\ \eta^\rho_{\ c}\gamma^a\gamma^c\gamma^b\partial_\mu\partial_\nu\partial_\rho\int d^4x'
\big[\langle\phi_+(x)\phi_+(x')\rangle-\langle\phi_+(x)\phi_-(x')\rangle\big]\hat{\psi}(x'). \notag
\end{align}
Here we have adopted the same approximation procedure with the scalar field case. 

Just like the scalar field theory case, we need to evaluate local terms to
estimate quantum IR effects due to soft gravitons: $\kappa^2H^2\log a(\tau)\gamma^a\partial_\mu\hat{\psi}(x)$.
For such a purpose,
we need to expand $\hat{\psi}(x')$ up to the first order
\begin{align}
\hat{\psi}(x')\to\hat{\psi}(x)-\partial_\alpha\hat{\psi}(x)\Delta x^\alpha. 
\label{Dexpansion}\end{align}
From (\ref{conformally1}) and (\ref{Dexpansion}), (\ref{Dint1}) is written as 
\begin{align}
\frac{\kappa^2H^2}{64\pi^4}\log a(\tau)&\eta^\rho_{\ c}\gamma^a\gamma^c\gamma^b\partial_\mu\partial_\nu\partial_\rho\int d^4x' \label{Dint2}\\
&\times\Big\{\big[\frac{1}{\Delta x_{++}^2}-\frac{1}{\Delta x_{+-}^2}\big]\hat{\psi}(x)
-\big[\frac{\Delta x^\alpha}{\Delta x_{++}^2}-\frac{\Delta x^\alpha}{\Delta x_{+-}^2}\big]\partial_\alpha\hat{\psi}(x)\Big\}. \notag
\end{align}
By substituting (\ref{C1}) and (\ref{C2}) to (\ref{Dint2}), (\ref{Dint1}) is evaluated as
\begin{align}
i\frac{\kappa^2H^2}{4\pi^2}\log a(\tau)\eta^\rho_{\ c}\gamma^a\gamma^c\gamma^b
\big\{-\frac{1}{4}(\delta_\mu^{\ 0}\delta_\nu^{\ 0}\partial_\rho+\delta_\mu^{\ 0}\delta_\rho^{\ 0}\partial_\nu+\delta_\nu^{\ 0}\delta_\rho^{\ 0}\partial_\mu)
+\frac{1}{2}\delta_\mu^{\ 0}\delta_\nu^{\ 0}\delta_\rho^{\ 0}\partial_0\big\}\hat{\psi}(x). 
\label{Dint3}\end{align}
From (\ref{gravityp}) and (\ref{Dint3}), the third term in (\ref{DEoM2}) is 
\begin{align}
\frac{\kappa^2H^2}{4\pi^2}\log a(\tau)\times i\big\{\frac{3}{16}\gamma^0\partial_0-\frac{5}{16}\gamma^i\partial_i\big\}\hat{\psi}(x). 
\label{DEoM4}\end{align}
From (\ref{DEoM3}) and (\ref{DEoM4}), the quantum equation of motion of a Dirac field
including the one-loop correction from soft gravitons is
\begin{align}
\big\{1+\frac{3\kappa^2H^2}{128\pi^2}\log a(\tau)\big\}
\times i\eta^\mu_{\ a}\gamma^a\partial_\mu\hat{\psi}(x). 
\label{Doverall}\end{align}
Just like a scalar field, the IR effect from gravitons to a Dirac field preserves the Lorentz invariance. 
It can be eliminated by the following time dependent wave function renormalization of a Dirac field:
\begin{align}
\psi\to Z_\psi\psi,\hspace{1em}Z_\psi\simeq 1-\frac{3\kappa^2H^2}{256\pi^2}\log a(\tau). 
\label{DZ}\end{align}
We also remark again that the IR logarithm can be absorbed into the wave function
renormalization factor even if we include a mass term
as a perturbation.

We summarize our investigations in this subsection. 
Inside the cosmological horizon, 
the IR effects of the gravitons at the one-loop level preserve the Lorentz invariance both in a free scalar and Dirac field theory. 
We suspect that this is the case beyond the one-loop level in the both scalar and Dirac field cases.
Although we think it is likely that soft graviton effects do not spoil Lorentz invariance of local physics,
so far we have only demonstrated it by explicit calculations.
We need to understand a mechanism to ensure it to all orders in perturbation theory.
Our analysis has shown that the IR effects manifest as the overall factors of the kinetic terms. 
In free field theories, they can be renormalized away by a time dependent wave function renormalization. 
With interaction, the wave function renormalization contributes to the renormalization 
of the coupling constants. 
We investigate the IR effects in $\phi^4$ theory and Yukawa theory in the next section. 

\subsection{Parametrization dependence}

We should compare our results to those in different parametrizations of the metric. 
As an example, the IR effects on a Dirac field have been investigated in the same gauge but in a different parametrization of the metric \cite{Miao2005(1),Miao2005(2),Miao2005(3)}. 
Here we make a parallel investigation in the parametrization: 
\begin{align}
g_{\mu\nu}=a^2(\tau)(\eta_{\mu\nu}+2\kappa\Phi(x)\eta_{\mu\nu}+\kappa\Psi_{\mu\nu}(x)),\hspace{1em}\eta^{\mu\nu}\Psi_{\mu\nu}=0. 
\label{Wpara}\end{align}
We have divided the fluctuation into the trace and traceless part 
to facilitate the comparison with our parametrization (\ref{para1})-(\ref{para3}). 
Furthermore, the authors of these papers adopt a different matter field redefinition from ours: 
\begin{align}
a^\frac{3}{2}\psi\to\psi. 
\label{WfrD}\end{align}
Since our calculations are rather heuristic in comparison to their dimensionally regulated and fully renormalized result,
it is eventually desirable to perform an analogous analysis.
Nevertheless our calculations are consistent with theirs with respect to how soft gravitons influence 
the local dynamics of the matter field at the sub-horizon scale as explained below.

Referring to Eq. (229) in \cite{Miao2005(1)}, 
the effective equation of the Dirac field is written as 
\begin{align}
i\eta^\mu_{\ a}\gamma^a\partial_\mu\hat{\psi}(x)-\int d^4x'\ c_{A}\Sigma_{+A}(x,x')\hat{\psi}(x')=0, 
\label{c_A}\end{align}
\begin{align}
\Sigma(x,x')=&\ \frac{i\kappa^2H^2}{64\pi^2}
\Big\{\frac{\log \big(a(\tau)a(\tau')\big)}{H^2a(\tau)a(\tau')}\partial^2\eta^\mu_{\ a}
+\frac{15}{2}\log \big(a(\tau)a(\tau')\big)\eta^\mu_{\ a} \label{self}\\
&\hspace{3.6em}-7\log \big(a(\tau)a(\tau')\big)(\eta^\mu_{\ a}-\delta^\mu_{\ 0}\delta^0_{\ a})\Big\}
\gamma^a\partial_\mu\delta^{(4)}(x-x') \notag\\
&+\frac{\kappa^2H^2}{256\pi^4}
\Big[\big\{\frac{1}{H^2a(\tau)a(\tau')}\partial^2\eta^\mu_{\ a}
+\frac{15}{2}\eta^\mu_{\ a}
-(\eta^\mu_{\ a}-\delta^\mu_{\ 0}\delta^0_{\ a})\big\}\gamma^a\partial_\mu\partial^2\frac{\log \mu^2\Delta x^2}{\Delta x^2} \notag\\
&\hspace{4.6em}-8(\eta^\mu_{\ a}-\delta^\mu_{\ 0}\delta^0_{\ a})
\gamma^a\partial_\mu\partial^2\frac{\log \frac{H^2}{4}\Delta x^2}{\Delta x^2}
+4\eta^\mu_{\ a}\gamma^a\partial_\mu\partial_i^2\frac{\log \frac{H^2}{4}\Delta x^2}{\Delta x^2} \notag\\
&\hspace{4.6em}+7\eta^\mu_{\ a}\gamma^a\partial_\mu\partial_i^2\frac{1}{\Delta x^2}\Big], \notag
\end{align}
where $c_A$ is identified as $c_\pm=\pm 1$ (the order of the double-sign corresponds).
From the self-energy (\ref{self}), the authors derive the solution of the effective field equation at the super-horizon scale. 
Specifically please see Eq. (32) in \cite{Miao2005(2)}, or Eq. (38) in \cite{Miao2005(3)}. 
In contrast, we investigate the off-shell effective field equation at the sub-horizon scale. 
We can recover the off-shell effective action from it up to a field independent function. 
One aim in this paper is to evaluate the wave function renormalization factor. 
We can uniquely determine it from the off-shell effective action by identifying the coefficient of the kinetic term, namely $\gamma^\mu p_\mu$ for a Dirac field for example 
(in the on-shell limit we need to evaluate the derivative with respect to $p_\mu$ since $\gamma^\mu p_\mu\hat{\psi}=0$). 
Note that Lorentz invariance must hold effectively in this procedure. 
In contrast to the flat space case, the result depends slowly on the conformal time. 
This IR logarithmic part grows large at late times. 
We evaluate only this part in our approximation procedure. 
Determining the full expression including the time independent part is beyond the scope of this paper.

In a similar way to (\ref{Dint1}) and (\ref{Dexpansion}), we can extract the local terms with the IR logarithms. 
The total of them is found as
\begin{align}
\big\{\eta^\mu_{\ a}-\frac{\kappa^2H^2}{16\pi^2}\log a(\tau)(\eta^{\mu}_{\ a}-\delta^\mu_{\ 0}\delta_{\ a}^0)\big\}\times i\gamma^a\partial_\mu\hat{\psi}(x). 
\label{WD}\end{align}
See Appendix \ref{A:C} for detailed calculations. 
The result (\ref{WD}) is different from our result (\ref{Doverall}). 
Furthermore it breaks the Lorentz invariance. 
The discrepancy originates just from the different choice of the parametrization of the metric and the matter field redefinition.  
That is, we can derive (\ref{WD}) from (\ref{Doverall}) by considering these differences within our approximation. 

Let us briefly explain the process of the derivation. 
We should note that the parametrization difference of the metric between (\ref{para1})-(\ref{para3}) and (\ref{Wpara}) 
emerges in the non-linear level: 
\begin{align}
\kappa w&=\kappa\Phi-\kappa^2\Phi^2-\frac{1}{16}\kappa^2\Psi_{\rho\sigma}\Psi^{\rho\sigma}+\cdots, \label{difference}\\
\kappa h_{\mu\nu}&=\kappa\Psi_{\mu\nu}-2\kappa^2\Phi\Psi_{\mu\nu}-\frac{1}{2}\kappa^2\Psi_\mu^{\ \rho}\Psi_{\rho\nu}+\frac{1}{8}\kappa^2\Psi_{\rho\sigma}\Psi^{\rho\sigma}\eta_{\mu\nu}+\cdots. \notag
\end{align} 
Then, as far as we adopt the same gauge
\begin{align}
F_\mu=\partial_\rho \Psi_\mu^{\ \rho}-2\partial_\mu \Phi+2\Psi_\mu^{\ \rho}\partial_\rho\log a+4\Phi\partial_\mu\log a, 
\end{align}
we have only to rename the field components to obtain the gravitational propagator in the parametrization (\ref{Wpara}): 
\begin{align}
w\to\Phi,\hspace{1em}h_{\mu\nu}\to\Psi_{\mu\nu}. 
\label{reword}\end{align}

If we do not consider the difference of the field redefinition, 
the parametrization difference of the metric (\ref{difference}) contributes only to the tadpole diagram: 
\begin{align}
\Delta(\delta\Gamma/\delta\hat{\psi})|_\text{metric}=
-i\frac{\kappa}{2}\langle (h^\mu_{\ a})_+(x)\rangle|_\text{NL}\gamma^a\partial_\mu\hat{\psi}(x), 
\label{PD1}\end{align}
where $\kappa\langle h_{\mu\nu}(x)\rangle|_\text{NL}$ is identified as
\begin{align}
\kappa\langle h_{\mu\nu}(x)\rangle|_\text{NL}
=-2\kappa^2\langle\Phi(x)\Psi_{\mu\nu}(x)\rangle
-\frac{1}{2}\kappa^2\langle\Psi_{\mu}^{\ \rho}(x)\Psi_{\rho\nu}(x)\rangle
+\frac{1}{8}\kappa^2\langle\Psi_{\rho\sigma}(x)\Psi^{\sigma\rho}(x)\rangle\eta_{\mu\nu}. 
\end{align}
From (\ref{gravityp}) and (\ref{reword}), 
(\ref{PD1}) is evaluated as 
\begin{align}
\Delta(\delta\Gamma/\delta\hat{\psi})|_\text{metric}\simeq
i\frac{\kappa^2H^2}{4\pi^2}\log a(\tau)\big\{-\frac{3}{8}\gamma^0\partial_0+\frac{1}{8}\gamma^i\partial_i\big\}\hat{\psi}(x). 
\label{PD2}\end{align}

In addition to the above, 
the field redefinition from (\ref{frD}) to (\ref{WfrD}) contributes to the quantum equation of motion as: 
\begin{align}
&\ \Delta(\delta\Gamma/\delta\hat{\psi})|_\text{field} \label{FD1}\\
=&\ \big\{3\kappa\langle w_+(x)\rangle|_\text{NL}\eta^\mu_{\ a}+\frac{9}{2}\kappa^2\langle w_+(x)w_+(x)\rangle\eta^\mu_{\ a}
-\frac{3}{2}\kappa^2\langle w_+(x)(h^\mu_{\ a})_+(x)\rangle\big\}
\gamma^a\partial_\mu\hat{\psi}(x) \notag\\
&-i\kappa^2\int d^4x'\ c_{AB} 
\big\{9\langle w_+(x)w_A(x')\rangle\eta^\mu_{\ a}\eta^\nu_{\ b}-\frac{3}{2}\langle w_+(x)(h^\nu_{\ b})_A(x')\rangle\eta^\mu_{\ a}\notag\\
&\hspace{8em}-\frac{3}{2}\langle (h^\mu_{\ a})_+(x)w_A(x')\rangle\eta^\nu_{\ b}\big\}\gamma^a\langle\partial_\mu\psi_+(x)\bar{\psi}_B(x')\rangle\gamma^b\partial_\nu'\hat{\psi}(x'), \notag
\end{align}
where $\kappa\langle w(x)\rangle|_\text{NL}$ originates in the parametrization difference of the metric (\ref{difference})
\begin{align}
\kappa\langle w(x)\rangle|_\text{NL}=-\kappa^2\langle\Phi^2(x)\rangle-\frac{1}{16}\kappa^2\langle\Psi_{\rho\sigma}(x)\Psi^{\rho\sigma}(x)\rangle. 
\end{align}
From (\ref{gravityp}), (\ref{Dint1}), (\ref{Dint3}) and (\ref{reword}), 
the following local term with the IR logarithm is reduced from (\ref{FD1}) 
\begin{align}
\Delta(\delta\Gamma/\delta\hat{\psi})|_\text{field}\simeq
i\frac{\kappa^2H^2}{4\pi^2}\log a(\tau)\big\{\frac{9}{32}\gamma^0\partial_0-\frac{15}{32}\gamma^i\partial_i\big\}\hat{\psi}(x). 
\label{FD2}\end{align}

By adding (\ref{PD2}) and (\ref{FD2}) to (\ref{Doverall}), we can derive (\ref{WD}). 
So we conclude that the discrepancy between (\ref{Doverall}) and (\ref{WD}) originates 
just from the different choice of the parametrization of the metric and the matter field redefinition. 
Although (\ref{WD}) breaks the Lorentz invariance, it can be eliminated by shifting the background metric.  
We will report the prescription to retain the Lorentz invariance elsewhere \cite{Parameter}. 

\section{$\phi^4$ theory and Yukawa theory}
\setcounter{equation}{0}

Here we investigate the IR effects from gravitons to interacting field theories. 
As specific examples, we  adopt $\phi^4$ theory and Yukawa theory. 
Since the coupling constants are dimensionless, 
$\sqrt{-g}$ can be absorbed by the field redefinition $\Omega\phi\to\phi,\ \Omega^\frac{3}{2}\psi\to\psi$
\begin{align}
\delta \mathcal{L}_4=-\frac{\lambda}{4!} \phi^4,  
\label{quartic}\end{align}
\begin{align}
\delta \mathcal{L}_Y=-g\phi\bar{\psi}\psi. 
\label{Yukawa}\end{align}
After the wave function renormalization (\ref{sZ}), (\ref{DZ}), 
the interaction terms are renormalized as 
\begin{align}
\delta \mathcal{L}_4=-\frac{\lambda}{4!} Z_\phi^4\phi^4,  
\label{quartic1}\end{align}
\begin{align}
\delta \mathcal{L}_Y=-g Z_\phi Z_\psi^2\phi\bar{\psi}\psi. 
\label{Yukawa1}\end{align} 

First, we investigate $\phi^4$ theory. 
Up to the one-loop level, the following nonlinear terms should be added to the left-hand side in (\ref{sEoM1})
\begin{align}
-\frac{\lambda}{6}Z_\phi^4\hat{\phi}^3(x)
&+\frac{\lambda\kappa^2}{2}\int d^4x'd^4x''\ c_{AB}c_{CD}\langle (h^{\mu\nu})_+(x)(h^{\rho\sigma})_A(x')\rangle\label{quartic2}\\
&\hspace{3em}\times\langle \partial_\mu\partial_\nu\phi_+(x)\phi_C(x'')\rangle\langle \partial_\rho'\partial_\sigma'\phi_B(x')\phi_D(x'')\rangle\hat{\phi}(x')\hat{\phi}^2(x'')\notag\\
&+\frac{\lambda\kappa^2}{2}\int d^4x'd^4x''\ c_{AB}c_{CD}\langle (h^{\mu\nu})_A(x')(h^{\rho\sigma})_C(x'')\rangle\notag\\
&\hspace{3em}\times\langle \phi_+(x)\partial_\mu'\partial_\nu'\phi_B(x')\rangle\langle \phi_+(x)\partial_\rho''\partial_\sigma''\phi_D(x'')\rangle\hat{\phi}(x)\hat{\phi}(x')\hat{\phi}(x''). \notag
\end{align}
Here we have performed the partial integration and neglected the differentiated gravitational field. 
The second and third terms denote the quantum correction to the vertex. 
The purpose in this section is to evaluate the effective coupling constant. 
To do that, we have only to extract the zeroth order of the classical fields
in the Taylor expansion of the relative coordinates
\begin{align}
\hat{\phi}(x')\hat{\phi}^2(x''),\ \hat{\phi}(x)\hat{\phi}(x')\hat{\phi}(x'')\to\hat{\phi}^3(x). 
\label{4expansion}\end{align}

In a similar way to (\ref{out1}) and (\ref{Dint1}), 
to evaluate these integrals up to $\kappa^2H^2\log a(\tau)$, 
we may adopt the following approximation
\begin{align}
\langle (h^{\mu\nu})_+(x)(h^{\rho\sigma})_+(x)\rangle\int d^4x'd^4x''. 
\label{out2}\end{align}
In this approximation, only the following local term contributes to the remaining integrals
\begin{align}
\langle\partial_\mu\partial_\nu\phi_+(x)\phi_+(x')\rangle
=-\langle\partial_\mu\phi_+(x)\partial_\nu'\phi_+(x')\rangle
\to -i\delta_\mu^{\ 0}\delta_\nu^{\ 0}\delta^{(4)}(x-x'). 
\label{local}\end{align}
As explained in the preceding subsection, the non-local term which we have neglected in (\ref{out2}) does not induce the dS symmetry breaking logarithm. 
As a result, (\ref{quartic2}) is evaluated as 
\begin{align}
&-\frac{\lambda}{6}Z_\phi^4\hat{\phi}^3(x)-\lambda\kappa^2\langle (h^{00})_+(x)(h^{00})_+(x)\rangle\hat{\phi}^3(x)\label{quartic3}\\
\simeq&-\frac{1}{6}\lambda\big\{1-\frac{21\kappa^2H^2}{16\pi^2}\log a(\tau)\big\}\hat{\phi}^3(x). \notag
\end{align}
In the second line, we have substituted (\ref{gravityp}), (\ref{minimallyc}) and (\ref{sZ}). 
(\ref{quartic3}) implies that the effective coupling constant in $\phi^4$ theory decreases with cosmic expansion under the influence of soft gravitons
\begin{align}
\lambda_\text{eff}\simeq\lambda\big\{1-\frac{21\kappa^2H^2}{16\pi^2}\log a(\tau)\big\}. 
\label{4ECC}\end{align}

Next, we investigate Yukawa theory. 
Up to the one-loop level, the following nonlinear terms should be added to the left-hand side in (\ref{DEoM1})
\begin{align}
-g Z_\phi Z_\psi^2\hat{\phi}(x)\hat{\psi}(x)
&+\frac{g\kappa^2}{4}\int d^4x'd^4x''\ c_{AB}c_{CD}\langle (h^\mu_{\ a})_+(x)(h^\nu_{\ b})_A(x')\rangle\label{Yukawa2}\\
&\hspace{3em}\times\langle \gamma^a\partial_\mu\psi_+(x)\bar{\psi}_C(x'')\rangle\langle \psi_D(x'')\partial_\nu'\bar{\psi}_B(x')\rangle\hat{\phi}(x'')\gamma^b\hat{\psi}(x')\notag\\
&+i\frac{g\kappa^2}{2}\int d^4x'd^4x''\ c_{AB}c_{CD}\langle (h^\mu_{\ a})_+(x)(h^{\rho\sigma})_A(x')\rangle\notag\\
&\hspace{3em}\times\langle \gamma^a\partial_\mu\psi_+(x)\bar{\psi}_C(x'')\rangle\langle \partial_\rho'\partial_\sigma'\phi_B(x')\phi_D(x'')\rangle\hat{\phi}(x')\hat{\psi}(x'')\notag\\
&-i\frac{g\kappa^2}{2}\int d^4x'd^4x''\ c_{AB}c_{CD}\langle (h^\mu_{\ a})_A(x')(h^{\rho\sigma})_C(x'')\rangle\notag\\
&\hspace{3em}\times\langle \psi_+(x)\partial_\mu'\bar{\psi}_B(x')\rangle\langle \phi_+(x)\partial_\rho''\partial_\sigma''\phi_D(x'')\rangle\hat{\phi}(x'')\gamma^a\hat{\psi}(x'). \notag
\end{align}
To evaluate the effective coupling constant, we have only to extract the zeroth order of the classical fields
in the Taylor expansion of the relative coordinates
\begin{align}
\hat{\phi}(x'')\hat{\psi}(x'),\ \hat{\phi}(x')\hat{\psi}(x'')\to\hat{\phi}(x)\hat{\psi}(x). 
\label{Yexpansion}\end{align}
By substituting $\langle\psi(x)\bar{\psi}(x')\rangle=i\eta^\nu_{\ c}\gamma^c\partial_\nu\langle\phi(x)\phi(x')\rangle$ 
and performing the parallel procedure with (\ref{out2})-(\ref{local}), (\ref{Yukawa2}) is evaluated as 
\begin{align}
&-g Z_\phi Z_\psi^2\hat{\phi}(x)\hat{\psi}(x)-g\cdot\frac{5\kappa^2}{4}\langle (h^{00})_+(x)(h^{00})_+(x)\rangle\hat{\phi}(x)\hat{\psi}(x)\label{Yukawa3}\\
\simeq&-g\big\{1-\frac{39\kappa^2H^2}{128\pi^2}\log a(\tau)\big\}\hat{\phi}(x)\hat{\psi}(x). \notag
\end{align}
We find that the effective coupling constant decreases with cosmic expansion
also in Yukawa theory:
\begin{align}
g_\text{eff}\simeq g\big\{1-\frac{39\kappa^2H^2}{128\pi^2}\log a(\tau)\big\}. 
\label{YECC}\end{align}

As seen in (\ref{4ECC}) and (\ref{YECC}), 
the gravitational fluctuations outside the cosmological horizon influence the physics inside the cosmological horizon at the one-loop level. 
By the power counting of the IR logarithms, 
the leading IR effect at the $n$-loop level is estimated as of order $(\kappa^2H^2\log a(\tau))^n$.
Since the perturbation theory is broken after $\kappa^2H^2\log a(\tau)\sim 1$, 
we need a nonperturbative method to understand its long term consequences.

Furthermore, we should emphasize that such IR effects on the dimensionless couplings can not be absorbed by the background metric. 
That is because the matter actions are independent of the conformal factor. 
As for the dynamics at the sub-horizon scale, the fact holds even when the quadratic matter action is non-conformally coupled. 
Thus the super-horizon gravitons certainly contribute to the sub-horizon dynamics, at least in the gauge (\ref{GF}). 
Of course we need to make sure that this is a physical effect. 
In order to clarify such an issue, we investigate the gauge dependence of these perturbative IR effects in the next section. 

\section{Gauge dependence}
\setcounter{equation}{0}

In the previous sections, we have investigated the IR effects with the gauge fixing term (\ref{GF}).  
It is important to investigate the gauge dependence of the obtained results. 
In this section, we adopt the following gauge fixing term with a parameter $\beta$: 
\begin{align}
\mathcal{L}_\text{GF}&=-\frac{1}{2}a^{D-2}\eta^{\mu\nu}F_\mu F_\nu, \label{beta}\\
F_\mu&=\beta\partial_\rho h_\mu^{\ \rho}-\beta(D-2)\partial_\mu w+\frac{1}{\beta}(D-2)h_\mu^{\ \rho}\partial_\rho\log a+\frac{2}{\beta}(D-2)w\partial_\mu\log a. \notag
\end{align}
This gauge fixing term coincides with (\ref{GF}) at $\beta=1$. 
For simplicity, we consider the case $|\beta^2-1|\ll 1$ since the deformation from (\ref{GF}) can be investigated perturbatively. 
The deformation of the action at $\mathcal{O}(\beta^2-1)$ is 
\begin{align}
\delta \mathcal{L}_{\beta^2-1}\simeq-\frac{1}{2}(\beta^2-1)a^2&\big[\ 
\eta^{\mu\nu}\partial_\mu h^{00}\partial_\nu h^{00}-3\partial_0 h^{00}\partial_0 h^{00}-\frac{5}{9}\partial_i h^{00}\partial_i h^{00}\label{beta1}\\
&-\frac{4}{3}\partial_i h^{00}\partial_k \tilde{h}^{ki}+\partial_k\tilde{h}^k_{\ i}\partial_l\tilde{h}^{li}\big]. 
\notag
\end{align}
Here we have set $D=4$ and neglected massless conformally coupled modes. 
In addition, we have ignored ghost fields since they do not couple to matter fields. 
To investigate the gauge dependence, we evaluate the correction to the gravitational propagator
from the additional term (\ref{beta1}). 

As seen in (\ref{out1}), (\ref{Dint1}) and (\ref{out2}), 
we only need to investigate the graviton propagator at the coincident point to evaluate the coefficient of $\kappa^2H^2\log a(\tau)$.    
From (\ref{gravityp}) and (\ref{beta1}), the IR logarithm can emerge in the following propagators
\begin{align}
\langle (h^{00})_+(x)(h^{00})_+(x)\rangle|_{\beta^2-1}
\simeq&-i(\beta^2-1)\int d^4x'\ a^2(\tau')c_{AB}\label{bint1}\\
&\times\big\{\ \eta^{\mu\nu}\langle(h^{00})_+(x)\partial_\mu'(h^{00})_A(x')\rangle\langle(h^{00})_+(x)\partial_\nu'(h^{00})_B(x')\rangle\notag\\
&\hspace{1.5em}-3\langle(h^{00})_+(x)\partial_0'(h^{00})_A(x')\rangle\langle(h^{00})_+(x)\partial_0'(h^{00})_B(x')\rangle\notag\\
&\hspace{1.5em}-\frac{5}{9}\langle(h^{00})_+(x)\partial_i'(h^{00})_A(x')\rangle\langle(h^{00})_+(x)\partial_i'(h^{00})_B(x')\rangle\big\}, \notag
\end{align}
\begin{align}
\langle (h^{00})_+(x)(\tilde{h}^{ij})_+(x)\rangle|_{\beta^2-1}
\simeq&-i(\beta^2-1)\int d^4x'\ a^2(\tau')c_{AB}\label{bint2}\\
&\times\big\{-\frac{2}{3}\langle(h^{00})_+(x)\partial_k'(h^{00})_A(x')\rangle\langle(\tilde{h}^{ij})_+(x)\partial_l'(\tilde{h}^{lk})_B(x')\rangle\big\}, \notag
\end{align}
\begin{align}
\langle (\tilde{h}^{ij})_+(x)(\tilde{h}^{kl})_+(x)\rangle|_{\beta^2-1}
\simeq&-i(\beta^2-1)\int d^4x'\ a^2(\tau')c_{AB}\label{bint3}\\
&\times\big\{\langle(\tilde{h}^{ij})_+(x)\partial_m'(\tilde{h}^m_{\ p})_A(x')\rangle\langle(\tilde{h}^{kl})_+(x)\partial_n'(\tilde{h}^{np})_B(x')\rangle\big\}. \notag 
\end{align}

To evaluate them, we need to perform the integration involving a massless and minimally coupled field 
\begin{align}
&-i(\beta^2-1)\int d^4x'\ a^2(\tau')c_{AB}\langle\varphi_+(x)\partial_\mu'\varphi_A(x')\rangle\langle\varphi_+(x)\partial_\nu'\varphi_B(x')\rangle. 
\label{bint0}\end{align}
In the integral, the following term of (\ref{minimally0}) contributes to the IR logarithm 
\begin{align}
\langle\varphi(x)\varphi(x')\rangle\simeq\frac{H^2}{8\pi^2}\big\{-\log y+\log a(\tau)a(\tau')\big\}=-\frac{H^2}{8\pi^2}\log H^2\Delta x^2. 
\label{minimallys}\end{align}
By substituting (\ref{minimallys}) and using the following identity  
\begin{align}
\int d^4x'\ \frac{1}{\tau'^2}\big[\frac{\Delta x_\mu\Delta x_\nu}{\Delta x^4_{++}}-\frac{\Delta x_\mu\Delta x_\nu}{\Delta x^4_{+-}}\big]
\simeq-4i\pi^2\log a(\tau)\big\{\delta_\mu^{\ 0}\delta_\nu^{\ 0}+\frac{1}{2}\eta_{\mu\nu}\big\}, 
\label{C4}\end{align}
the integral (\ref{bint0}) is evaluated as 
\begin{align}
-(\beta^2-1)\frac{H^2}{4\pi^2}\log a(\tau)\big\{\delta_\mu^{\ 0}\delta_\nu^{\ 0}+\frac{1}{2}\eta_{\mu\nu}\big\}. 
\label{bint0result}\end{align} 
We explain how to derive the identity (\ref{C4}) in Appendix \ref{A:A}. 

From (\ref{bint0result}), (\ref{bint1})-(\ref{bint3}) are 
\begin{align}
\langle (h^{00})_+(x)(h^{00})_+(x)\rangle|_{\beta^2-1}&\simeq-(\beta^2-1)\times-\frac{3}{4}\frac{H^2}{4\pi^2}\log a(\tau), \label{bgravityp}\\
\langle (h^{00})_+(x)(\tilde{h}^{ij})_+(x)\rangle|_{\beta^2-1}&\simeq 0, \notag\\
\langle (\tilde{h}^{ij})_+(x)(\tilde{h}^{kl})_+(x)\rangle|_{\beta^2-1}&\simeq-(\beta^2-1)\times(\delta^{ik}\delta^{jl}+\delta^{il}\delta^{jk}-\frac{2}{3}\delta^{ij}\delta^{kl})\frac{H^2}{4\pi^2}\log a(\tau). \notag
\end{align}
We should note that the deformation of the propagator is proportional to the original one. 
As a result, in the deformed gauge (\ref{beta}), 
we have only to replace the gravitational propagator as follows 
\begin{align}
\langle (h^{\mu\nu})_+(x)(h^{\rho\sigma})_+(x)\rangle
\to\big\{1-(\beta^2-1)\big\}\langle (h^{\mu\nu})_+(x)(h^{\rho\sigma})_+(x)\rangle. 
\label{bgravityp'}\end{align}
From this fact, we can conclude that the Lorentz invariance is preserved for a continuous $\beta$ 
\begin{align}
Z_\phi&\simeq 1-(2-\beta^2)\frac{3\kappa^2H^2}{64\pi^2}\log a(\tau), \label{betaZ}\\
Z_\psi&\simeq 1-(2-\beta^2)\frac{3\kappa^2H^2}{256\pi^2}\log a(\tau). 
\end{align} 
However the effective couplings are found to be gauge dependent.
\begin{align}
\lambda_\text{eff}&\simeq\lambda\big\{1-(2-\beta^2)\frac{21\kappa^2H^2}{16\pi^2}\log a(\tau)\big\}, \label{betaECC}\\ 
g_\text{eff}&\simeq g\big\{1-(2-\beta^2)\frac{39\kappa^2H^2}{128\pi^2}\log a(\tau)\big\}. \notag
\end{align}

Here we summarize our findings as follows. In the original gauge
the dimensionless coupling constants $\lambda$ and $g$ 
are screened by soft gravitons and decrease with time in dS space.
They acquire nontrivial scaling exponents in dS space as
\begin{align}
\lambda_\text{eff}=\lambda f(t)^\frac{21}{4},~~~g_\text{eff}=g f(t)^\frac{39}{32}
\end{align}
where $f(t)=1-\frac{\kappa^2 H^2}{4\pi^2}Ht$. 
In order to confirm that this is a physical effect, 
we have examined the gauge parameter dependence of these results 
using one parameter family of the graviton propagator.
As it turns out $f(t)$ depends on a gauge parameter $\beta$ as
$f(t)=1-(2-\beta^2)\frac{\kappa^2 H^2}{4\pi^2}Ht$.
Nevertheless we observe that the following relative scaling relation is independent of a gauge parameter 
\begin{align}
g_\text{eff}\sim (\lambda_\text{eff})^\frac{13}{56}. 
\label{result}\end{align}

We point out that the situation here is analogous to the scaling exponents of the operators in two
dimensional quantum gravity. 
The scaling exponents of the individual operators are gauge dependent.
However the relative scaling exponents are gauge invariant \cite{KN,KKN}.
It is because there is no unique way to specify the scale there.
Analogously  in our case there is no unique way to specify the time as it depends on an observer.
Here a sensible strategy is to pick a particular coupling and use its time evolution as
a physical time. In this setting the relative scaling exponents measure the speed of
the time evolution of the couplings in terms of a physical time.
Although the choice of time is not unique,
the relative scaling exponents are gauge independent and well defined.
We still need to check the validity of this picture against large deformations
of a gauge parameter.

\section{Conclusion}
\setcounter{equation}{0}

The inflation theory postulates that almost scale invariant density perturbation 
is generated by quantum fluctuations in a dS type space.
The detection of almost scale invariant gravitational fluctuations is a crucial test
of the inflation theory since it universally generates them.
Thus the propagator of gravitons contains dS symmetry breaking IR logarithms in dS space. 

In this paper we have investigated its physical implications on microscopic physics.
Namely we have investigated the effects of super-horizon modes of gravitons on the dynamics of the sub-horizon modes of matter fields. 
By evaluating the kinetic terms of scalar and Dirac fields up to the one-loop level, 
we have found that the IR effects from gravitons preserve the Lorentz invariance. 
In particular the velocity of massless particles remains universal irrespective of the spins. 
Then soft graviton effects in free field theories can be absorbed into the wave function renormalization of scalar and Dirac fields. 

It is important to prove our results beyond the approximations we have employed. 
We have checked our results against an exact calculation for a diagram in Appendix B. Another check is a field theoretic consistency. 
The wave function renormalization factor in Dirac field theory can also be extracted from the previously known self-energy (\ref{self}) in the off-shell regime. 
The resultant quantum equation (\ref{WD}) is consistent with ours (\ref{Doverall}) although they are different. 
It is due to different metric parametrizations and the choice of fundamental field in both calculations. 
We will report that we can eliminate the parametrization dependence by shifting the background metric \cite{Parameter}. 

In the interacting field theory with dimensionless couplings, we have found that the couplings become time dependent and decrease with time. 
It is very important to check whether our findings are not gauge artifacts. 
The Lorentz invariance is preserved even when the gauge parameter is slightly deformed.  
In fact, the screening of each coupling is gauge dependent. 
That makes sense since there is no unique way to specify the time in quantum gravity. 
Of course, we need to find a physical interpretation with gauge invariant quantities for the screening of couplings.  
We have identified candidates of physical observables in the relative scaling exponents which are invariant under infinitesimal gauge change. 
They measure the relative evolution speed of the couplings among matter fields in terms of a physical time. 
Up to the one-loop level, they are just numbers like $13/56$ in (\ref{result}) which do not depend on initial time.  

It is imperative to understand why soft gravitons diminish dimensionless couplings with time. 
We admit here that we have not found a persuasive mechanism. 
We merely describe the following observation. 
The dimensionless couplings couples to marginal operators in 4 dimensions. 
They become irrelevant operators in higher dimensions. 
It might imply IR quantum fluctuations of dS space increase the effective dimension of space-time. 
If so, this kind of effect may decrease dimensionless couplings with cosmic evolution. 
We may draw an analogy with 2 dimensional quantum gravity again. 
There the quantum fluctuation of geometry not only increases the dimension of the local operators 
but also the (fractal) dimension of space-time \cite{KN,KKN}. 
We hope to see whether such an observation leads to an understanding of these symmetry breaking effects. 

The results obtained in this paper are the one-loop effects.  
Since the IR effects at each loop level manifest as polynomials in $\log a(\tau)$, perturbation theories are broken after enough time passed. 
In other words, the obtained results describe physics at the initial stage $\kappa^2H^2\log a(\tau)\ll 1$. 
To investigate the eventual contributions to physical quantities, we need to evaluate the IR effects nonperturbatively. 
Although the IR effects from specific matter fields have been investigated nonperturbatively \cite{Starobinsky1994,Woodard2005,Woodard2006,Prokopec2007,Kitamoto2011}, 
the nonperturbative approach for the IR effects from gravitons is an open issue. 
   
Our Universe is dominated by dark energy. Thus our results imply that the couplings of the standard model become time dependent. 
Since the coefficients of IR logarithms are of $O(\kappa^2 H^2)\sim 10^{-120}$, it is not observable now. 
However it is relevant to the ultimate fate of the Universe as their effect grows linearly with cosmic time like Hawking radiation. 
This effect is much larger in the inflationary era as $\kappa^2 H^2$ could be as large as $10^{-10}$. 

Since the cosmological constant is a function of the couplings of the microscopic theory, 
it may acquire time dependence if the couplings evolve with time. 
So we need to investigate such effects to understand possible time dependence of the cosmological constant in addition to pure matter and gravity contributions separately. 
Let us follow the change of the couplings and cosmological constant under time evolution. 
We note here that the vanishing cosmological constant is a fixed point of the evolution where the couplings become constant. 
It seems conceivable that such a self-tuning mechanism may play an important role in the cosmological constant problem. 

\section*{Acknowledgment}
This work is supported in part by the Grant-in-Aid for Scientific Research
from the Ministry of Education, Science and Culture of Japan. 
We thank our referees for asking question which helps us to clarify many important issues. 

\appendix
\section{Derivation of (\ref{C1})-(\ref{C3}) and (\ref{C4})}\label{A:A}
\setcounter{equation}{0}

Here we explain how to derive the identities (\ref{C1})-(\ref{C3}) and (\ref{C4}). 
Each integrand in its left-hand side is written as
\begin{align}
\frac{1}{\Delta x^2}&=\frac{1}{4}\partial^2\log H^2\Delta x^2, 
\end{align}
\begin{align}
\frac{\Delta x_\alpha}{\Delta x^2}&=\frac{1}{2}\partial_\alpha\log H^2\Delta x^2, 
\end{align}
\begin{align}
\frac{\Delta x_\alpha\Delta x_\beta}{\Delta x^2}&=\frac{1}{2}\partial_\alpha(\Delta x_\beta\log H^2\Delta x^2)-\frac{1}{2}\eta_{\alpha\beta}\log H^2\Delta x^2, 
\end{align}
\begin{align}
\frac{\Delta x_\alpha\Delta x_\beta}{\Delta x^4}=-\frac{1}{4}\big\{\partial_\alpha\partial_\beta-\frac{1}{2}\eta_{\alpha\beta}\partial^2\big\}\log H^2\Delta x^2, 
\end{align}
where we abbreviate the indexes $++,\ +-$ because the above identities work in both cases. 
By using them and leaving differential operators out of the integrals, the left-hand sides in (\ref{C1})-(\ref{C3}) and (\ref{C4}) are
\begin{align}
&\partial_\alpha\partial_\beta\int d^4x'\big[\frac{1}{\Delta x_{++}^2}-\frac{1}{\Delta x_{+-}^2}\big] \label{C11}\\
=&-\frac{1}{4}\delta_\alpha^{\ 0}\delta_\beta^{\ 0}\partial_0^4\int d^4x'\big[\log H^2\Delta x_{++}^2-\log H^2\Delta x_{+-}^2\big], \notag
\end{align} 
\begin{align}
&\partial_\beta\partial_\gamma\partial_\delta\int d^4x'\big[\frac{\Delta x_\alpha}{\Delta x_{++}^2}-\frac{\Delta x_\alpha}{\Delta x_{+-}^2}\big] \label{C21}\\
=&\ \frac{1}{2}\delta_\alpha^{\ 0}\delta_\beta^{\ 0}\delta_\gamma^{\ 0}\delta_\delta^{\ 0}\partial_0^4\int d^4x'\big[\log H^2\Delta x_{++}^2-\log H^2\Delta x_{+-}^2\big], \notag
\end{align}
\begin{align}
&\partial_\gamma\partial_\delta\partial_\varepsilon\partial_\delta
\int d^4x'\big[\frac{\Delta x_\alpha\Delta x_\beta}{\Delta x_{++}^2}-\frac{\Delta x_\alpha\Delta x_\beta}{\Delta x_{+-}^2}\big] \label{C31}\\
=&-\frac{1}{2}\delta_\alpha^{\ 0}\delta_\beta^{\ 0}\delta_\gamma^{\ 0}\delta_\delta^{\ 0}\delta_\varepsilon^{\ 0}\delta_\eta^{\ 0}
\partial_0^5\int d^4x'\ \Delta\tau\big[\log H^2\Delta x_{++}^2-\log H^2\Delta x_{+-}^2\big] \notag\\
&-\frac{1}{2}\eta_{\alpha\beta}\delta_\gamma^{\ 0}\delta_\delta^{\ 0}\delta_\varepsilon^{\ 0}\delta_\eta^{\ 0}
\partial_0^4\int d^4x'\big[\log H^2\Delta x_{++}^2-\log H^2\Delta x_{+-}^2\big], \notag
\end{align}
\begin{align}
&\int d^4x'\ \frac{1}{\tau'^2}\big[\frac{\Delta x_\alpha\Delta x_\beta}{\Delta x_{++}^4}-\frac{\Delta x_\alpha\Delta x_\beta}{\Delta x_{+-}^4}\big] \label{C41}\\
=&-\frac{1}{4}\big\{\delta_\alpha^{\ 0}\delta_{\beta}^{\ 0}+\frac{1}{2}\eta_{\alpha\beta}\big\}\partial_0^2\int d^4x'\ \frac{1}{\tau'^2}\big[\log H^2\Delta x_{++}^2-\log H^2\Delta x_{+-}^2\big]. \notag
\end{align}
Note that the differential operator outside the integral is equal to the time derivative $\partial_\alpha\to\delta_\alpha^{\ 0}\partial_0$. 
Furthermore, we have replaced $\Delta x_\alpha$ to $-\delta_\alpha^{\ 0}\Delta\tau$ in (\ref{C31}).  

As a concrete example, we calculate the following integral 
\begin{align}
\partial_0^4\int d^4x'\big[\log H^2\Delta x_{++}^2-\log H^2\Delta x_{+-}^2\big]. 
\label{int1}\end{align}
From (\ref{pm}), the logarithm with each index $++,\ +-$ is
\begin{align}
\log H^2\Delta x_{++}^2&=\log H^2|\Delta\tau^2-r^2|+i\pi\theta(\Delta\tau^2-r^2), \label{log}\\
\log H^2\Delta x_{+-}^2&=\log H^2|\Delta\tau^2-r^2|-i\pi\theta(\Delta\tau^2-r^2)\big\{\theta(\Delta \tau)-\theta(-\Delta\tau)\big\}, \notag
\end{align}
where $r^2\equiv({\bf x}-{\bf x}')^2$. 
By substituting (\ref{log}), (\ref{int1}) is
\begin{align}
\partial_0^4\int^\tau_{\tau_i} d\tau'\int^{\Delta\tau}_0 4\pi r^2dr\ 2i\pi 
=16i\pi^2\partial_0\int^\tau_{\tau_i} d\tau' =16i\pi^2. 
\label{int1'}\end{align}
The initial time is identified as $\tau_i=-1/H$ since we set the IR cut-off as $\epsilon_0=H$. 
In a similar way, the other integrals are evaluated as
\begin{align}
\partial_0^5\int d^4x'\ \Delta\tau\big[\log H^2\Delta x_{++}^2-\log H^2\Delta x_{+-}^2\big]
=64i\pi^2\partial_0\int^\tau_{\tau_i} d\tau' =64i\pi^2, 
\label{int2}\end{align}
\begin{align}
\partial_0^2\int d^4x'\ \frac{1}{\tau'^2}\big[\log H^2\Delta x_{++}^2-\log H^2\Delta x_{+-}^2\big] 
&=16i\pi^2\int^\tau_{\tau_i} d\tau'\ \frac{\Delta\tau}{\tau'^2} \label{int3}\\
&\simeq 16i\pi^2\log a(\tau). \notag
\end{align}
It should be noted that the integrals (\ref{int1'}), (\ref{int2}) do not depend on the lower bound  $\tau_i$. 
By substituting (\ref{int1'})-(\ref{int3}) to (\ref{C11})-(\ref{C41}), we obtain the desired identities
\begin{align}
\partial_\alpha\partial_\beta\int d^4x'\big[\frac{1}{\Delta x_{++}^2}-\frac{1}{\Delta x_{+-}^2}\big]
=-4i\pi^2\delta_\alpha^{\ 0}\delta_\beta^{\ 0}, 
\label{C1'}\end{align}
\begin{align}
\partial_\beta\partial_\gamma\partial_\delta\int d^4x'\big[\frac{\Delta x_\alpha}{\Delta x_{++}^2}-\frac{\Delta x_\alpha}{\Delta x_{+-}^2}\big]
= 8i\pi^2\delta_\alpha^{\ 0}\delta_\beta^{\ 0}\delta_\gamma^{\ 0}\delta_\delta^{\ 0}, 
\label{C2'}\end{align}
\begin{align}
&\partial_\gamma\partial_\delta\partial_\varepsilon\partial_\eta\int d^4x'\big[\frac{\Delta x_\alpha\Delta x_\beta}{\Delta x_{++}^2}-\frac{\Delta x_\alpha\Delta x_\beta}{\Delta x_{+-}^2}\big]\label{C3'}\\
=&-32i\pi^2\delta_\alpha^{\ 0}\delta_\beta^{\ 0}\delta_\gamma^{\ 0}\delta_\delta^{\ 0}\delta_\varepsilon^{\ 0}\delta_\eta^{\ 0}
-8i\pi^2\eta_{\alpha\beta}\delta_\gamma^{\ 0}\delta_\delta^{\ 0}\delta_\varepsilon^{\ 0}\delta_\eta^{\ 0}, \notag
\end{align}
\begin{align}
\int d^4x'\ \frac{1}{\tau'^2}\big[\frac{\Delta x_\mu\Delta x_\nu}{\Delta x^4_{++}}-\frac{\Delta x_\mu\Delta x_\nu}{\Delta x^4_{+-}}\big]
\simeq-4i\pi^2\log a(\tau)\big\{\delta_\mu^{\ 0}\delta_\nu^{\ 0}+\frac{1}{2}\eta_{\mu\nu}\big\}.  
\label{C4'}\end{align}

\section{Exact evaluation for (\ref{sint1})}\label{A:B}
\setcounter{equation}{0}

In this appendix, we evaluate (\ref{sint1}) without using the approximation method (\ref{minimallya})-(\ref{out1}). 
From (\ref{sint1}), the following integrals contribute to the second derivative of $\hat{\phi}(x)$: 
\begin{align}
&i\kappa^2\partial_\mu\int d^4x'\ c_{AB}\partial_\sigma'\big\{\langle \varphi_+(x)\varphi_A(x')\rangle\langle\partial_\nu\phi_+(x)\partial_\rho'\phi_B(x')\rangle\big\}\hat{\phi}(x')\label{Is0}\\
\to &\ (\sum_n I^n_{\ \mu\nu\rho\sigma\alpha\beta})\times\partial^\alpha\partial^\beta\hat{\phi}(x), \notag
\end{align}
\begin{align}
I^1_{\ \mu\nu\rho\sigma\alpha\beta}
&=i\kappa^2\eta_{\mu\beta}\int d^Dx'\ c_{AB}\partial_\sigma'\big\{\Delta x_\alpha\langle \varphi_+(x)\varphi_A(x')\rangle\langle\partial_\nu\partial_\rho\phi_+(x)\phi_B(x')\rangle\big\}, \label{Is1}\\
I^2_{\ \mu\nu\rho\sigma\alpha\beta}
&=i\kappa^2\eta_{\mu\beta}\eta_{\sigma\alpha}\partial_\rho\int d^Dx'\ c_{AB}\big\{\langle \varphi_+(x)\varphi_A(x')\rangle\langle\partial_\nu\phi_+(x)\phi_B(x')\rangle\big\}, \notag\\
I^3_{\ \mu\nu\rho\sigma\alpha\beta}
&=-i\kappa^2\eta_{\mu\beta}\eta_{\sigma\alpha}\int d^Dx'\ c_{AB}\big\{\langle \partial_\rho\varphi_+(x)\varphi_A(x')\rangle\langle\partial_\nu\phi_+(x)\phi_B(x')\rangle\big\}, \notag\\
I^4_{\ \mu\nu\rho\sigma\alpha\beta}
&=-i\frac{\kappa^2}{2}\partial_\mu\int d^Dx'\ c_{AB}\partial_\sigma'
\big\{\Delta x_\alpha\Delta x_\beta\langle\varphi_+(x)\varphi_A(x')\rangle\langle\partial_\nu\partial_\rho\phi_+(x)\phi_B(x')\rangle\big\}, \notag\\
I^5_{\ \mu\nu\rho\sigma\alpha\beta}
&=-i\kappa^2\eta_{\sigma\beta}\partial_\mu\int d^Dx'\ c_{AB}\big\{\Delta x_\alpha\langle\varphi_+(x)\varphi_A(x')\rangle\langle\partial_\nu\partial_\rho\phi_+(x)\phi_B(x')\rangle\big\}. \notag
\end{align}
To evaluate these integrals, we adopt the dimensional regularization $D=4-\varepsilon$ \cite{Miao2008,Miao2010}: 
\begin{align}
\langle\varphi(x)\varphi(x')\rangle=\frac{H^{D-2}}{4\pi^\frac{D}{2}}\big\{\frac{\Gamma(\frac{D}{2}-1)}{y^{\frac{D}{2}-1}}-\frac{1}{2}\log H^2\Delta x^2+1-\gamma\big\}, 
\label{minimallyD}\end{align}
\begin{align}
\langle\phi(x)\phi(x')\rangle=\frac{\Gamma(\frac{D}{2}-1)}{4\pi^\frac{D}{2}}\frac{1}{\Delta x^{D-2}}.  
\label{conformallyD}\end{align}
We have set $\varepsilon=0$ in the second and third terms of (\ref{minimallyD}).  

For example, we calculate $I^2_{\ \mu\nu\rho\sigma\alpha\beta}$. 
From (\ref{minimallyD}) and (\ref{conformallyD}), $I^2_{\ \mu\nu\rho\sigma\alpha\beta}$ is written as
\begin{align}
I^2_{\ \mu\nu\rho\sigma\alpha\beta}=&-i\eta_{\mu\beta}\eta_{\sigma\alpha}\frac{\Gamma(\frac{D}{2})\kappa^2H^{D-2}}{8\pi^D}\partial_\rho\int d^Dx'\ c_A\\
&\times\big\{\frac{\Gamma(1-\frac{\varepsilon}{2})}{H^{2-\varepsilon}}a^{-1+\frac{\varepsilon}{2}}(\tau)a^{-1+\frac{\varepsilon}{2}}(\tau')\frac{\Delta x_\nu}{\Delta x^{6-2\varepsilon}_{+A}}
-\frac{1}{2}\frac{\Delta x_\nu}{\Delta x^4_{+A}}\log H^2\Delta x^2_{+A}+(1-\gamma)\frac{\Delta x_\nu}{\Delta x^4_{+A}}\big\}. \notag
\end{align}
Here $c_A$ is defined in the same way as in Eq. (\ref{c_A}). 
The second term of the integrand comes from the dS broken part of (\ref{minimallyD}) 
and the other terms come from the dS invariant part. 
We show that only the dS broken part contributes to the coefficient of $\log a(\tau)$ by explicit calculation. 

By substituting the following identities, 
\begin{align}
\frac{\Delta x_\nu}{\Delta x^{6-2\varepsilon}}=\frac{1}{(4-2\varepsilon)(2-2\varepsilon)}\partial_\nu\frac{\partial^2}{\varepsilon}\frac{1}{\Delta x^{2-2\varepsilon}}, 
\end{align}
\begin{align}
\frac{\Delta x_\nu}{\Delta x^4}\log H^2\Delta x^2=-\frac{1}{16}\partial_\nu\partial^2\log^2 H^2\Delta x^2, 
\end{align}
\begin{align}
\frac{\Delta x_\nu}{\Delta x^4}=-\frac{1}{8}\partial_\nu\partial^2\log H^2\Delta x^2, 
\end{align}
each integral is written as 
\begin{align}
&\partial_\rho\int d^Dx'\ a^{-1+\frac{\varepsilon}{2}}(\tau)a^{-1+\frac{\varepsilon}{2}}(\tau')
\big[\frac{\Delta x_\nu}{\Delta x^{6-2\varepsilon}_{++}}-\frac{\Delta x_\nu}{\Delta x^{6-2\varepsilon}_{+-}}\big]\label{I21}\\
=&\ \frac{\delta_\nu^{\ 0}\delta_\rho^{\ 0}}{(4-2\varepsilon)(2-2\varepsilon)}\partial_0\Big\{
a^{-1+\frac{\varepsilon}{2}}(\tau)\partial_0\int d^Dx'\ a^{-1+\frac{\varepsilon}{2}}(\tau')\frac{\partial^2}{\varepsilon}\big[\frac{1}{\Delta x^{2-2\varepsilon}_{++}}-\frac{1}{\Delta x^{2-2\varepsilon}_{+-}}\big]
\Big\}, \notag
\end{align}
\begin{align}
&\partial_\rho\int d^4x'\ \big[\frac{\Delta x_\nu}{\Delta x^4_{++}}\log H^2\Delta x^2_{++}-
\frac{\Delta x_\nu}{\Delta x^4_{+-}}\log H^2\Delta x^2_{+-}\big]\label{I22}\\
=&\ \frac{1}{16}\delta_\nu^{\ 0}\delta_\rho^{\ 0}\partial^4_0\int d^4x'\ \big[\log^2 H^2\Delta x^2_{++}-\log^2 H^2\Delta x^2_{+-}\big], \notag
\end{align}
\begin{align}
&\partial_\rho\int d^4x'\ \big[\frac{\Delta x_\nu}{\Delta x^4_{++}}-\frac{\Delta x_\nu}{\Delta x^4_{+-}}\big]\label{I23}\\
=&\ \frac{1}{8}\delta_\nu^{\ 0}\delta_\rho^{\ 0}\partial^4_0\int d^4x'\ \big[\log H^2\Delta x^2_{++}-\log H^2\Delta x^2_{+-}\big]. \notag
\end{align}
From (\ref{log}), (\ref{I23}) is evaluated as
\begin{align}
2i\pi^2\delta_\nu^{\ 0}\delta_\rho^{\ 0}\partial_0\int^\tau_{\tau_i} d\tau'
\simeq 0\times\log a(\tau). 
\label{I23'}\end{align}

In a similar way, (\ref{I22}) is evaluated as 
\begin{align}
&\ 4i\pi^2\delta_\nu^{\ 0}\delta_\rho^{\ 0}\partial_0\int^\tau_{\tau_i}d\tau'\ \big\{\log 2H\Delta\tau+\frac{1}{2}\big\}\label{I22'}\\
=&\ 4i\pi^2\delta_\nu^{\ 0}\delta_\rho^{\ 0}\partial_0\int^\tau_{\tau_i}d\tau'\ \big\{\log(-H\tau')+\frac{1}{2}+\log 2(1-\frac{\tau}{\tau'})\big\}. \notag
\end{align}
The first two integrals lead to the following terms 
\begin{align}
4i\pi^2\delta_\nu^{\ 0}\delta_\rho^{\ 0}\big\{-\log a(\tau)+\frac{1}{2}\big\}. 
\label{L}\end{align}
We should note that these terms come from the derivatives of the upper bound $\tau$. 
It indicates that they describe the local dynamics. 
In contrast, the remaining integral describes the non-local contribution.  
The time integral seems to induce the dS symmetry breaking logarithm: 
\begin{align}
4i\pi^2\delta_\nu^{\ 0}\delta_\rho^{\ 0}\partial_0\int^\tau_{\tau_i}d\tau'\ \log 2(1-\frac{\tau}{\tau'})
\simeq -4i\pi^2\delta_\nu^{\ 0}\delta_\rho^{\ 0}\partial_0\int^\tau_{\tau_i}d\tau'\ \frac{\tau}{\tau'}. 
\label{NL}\end{align}
Thus the approximation to neglect differentiated gravitational fields and the dS invariant part of graviton propagators seems to fail at first sight. 
Let us recall that we investigate the off-shell effective equation in this paper. 
It is equivalent to substituting the off-shell classical field $\hat{\phi}(x')\propto e^{ip_\mu x'^{\mu}},\ p_\mu p^\mu\not=0$ into the integral. 
In this case, the integral over the negatively large conformal time is suppressed by an inverse of the virtuality: 
\begin{align}
\int^\tau_{-1/\sqrt{-p_\mu p^\mu}}\frac{d\tau'}{\tau'}=\log (-\sqrt{p_\mu p^\mu}\tau). 
\label{bound}\end{align}
Note that the logarithm is invariant under the scaling transformation (\ref{scaling}). 
In other words, such a logarithm is time independent when it is expressed by the physical momentum scale $-H\tau\sqrt{p_\mu p^\mu}$. 
That is why we have neglected the non-local terms. 
In order to probe local dynamics of matter fields, we can perform experiments in the laboratory. 
We thus hold the physical momentum fixed in such a situation. 
Considering (\ref{I22'})-(\ref{bound}), (\ref{I22}) induces the following dS broken term 
\begin{align}
4i\pi^2\delta_\nu^{\ 0}\delta_\rho^{\ 0}\partial_0\int^\tau_{\tau_i}d\tau'\ \big\{\log 2H\Delta\tau+\frac{1}{2}\big\}
\simeq -4i\pi^2\delta_\nu^{\ 0}\delta_\rho^{\ 0}\log a(\tau). 
\label{I22''}\end{align}
The scaling law (\ref{general}) is correct only when we consider the replacement of the lower bound as in (\ref{bound}). 
Then the approximation adopted in this paper is valid to evaluate the dS symmetry breaking.    

To evaluate (\ref{I21}), we need to extract the $1/\varepsilon$ part as follows: 
\begin{align}
\partial^2\frac{1}{\Delta x_{++}^{2-\varepsilon}}=\frac{4i\pi^{2-\frac{\varepsilon}{2}}}{\Gamma(1-\frac{\varepsilon}{2})}\delta^{(D)}(\Delta x),\hspace{1em}
\partial^2\frac{1}{\Delta x_{+-}^{2-\varepsilon}}=0
\end{align}
\begin{align}
\frac{\partial^2}{\varepsilon}\frac{1}{\Delta x_{++}^{2-2\varepsilon}}
&=\frac{\partial^2}{\varepsilon}\big\{\frac{1}{\Delta x_{++}^{2-2\varepsilon}}-\frac{\mu^{-\varepsilon}}{\Delta x_{++}^{2-\varepsilon}}\big\}
+\frac{4i\pi^{2-\frac{\varepsilon}{2}}\mu^{-\varepsilon}}{\Gamma(1-\frac{\varepsilon}{2})\varepsilon}\delta^{D}(\Delta x) \label{UV}\\
&=\frac{\partial^2}{2}\big\{\frac{\log \mu^2\Delta x^2_{++}}{\Delta x^2_{++}}\big\}
+\frac{4i\pi^{2-\frac{\varepsilon}{2}}\mu^{-\varepsilon}}{\Gamma(1-\frac{\varepsilon}{2})\varepsilon}\delta^{D}(\Delta x), \notag\\
\frac{\partial^2}{\varepsilon}\frac{1}{\Delta x_{+-}^{2-2\varepsilon}}
&=\frac{\partial^2}{\varepsilon}\big\{\frac{1}{\Delta x_{+-}^{2-2\varepsilon}}-\frac{\mu^{-\varepsilon}}{\Delta x_{++}^{2-\varepsilon}}\big\}\notag\\
&=\frac{\partial^2}{2}\big\{\frac{\log \mu^2\Delta x^2_{+-}}{\Delta x^2_{+-}}\big\}, \notag
\end{align}
where we introduce the mass parameter $\mu$ to correct the dimension. 
By using (\ref{UV}) and the following identity, 
\begin{align}
\frac{\log \mu^2\Delta x^2}{\Delta x^2}=\frac{1}{8}\partial^2\big\{\log^2\mu^2\Delta x^2-2\log \mu^2\Delta x^2\big\}, 
\end{align}
(\ref{I21}) is evaluated as
\begin{align}
&\frac{4i\pi^2\delta_\nu^{\ 0}\delta_\rho^{\ 0}}{(4-2\varepsilon)(2-2\varepsilon)}\partial_0\Big\{a^{-1+\frac{\varepsilon}{2}}(\tau)\partial_0
\big\{\frac{\pi^{-\frac{\varepsilon}{2}}\mu^{-\varepsilon}a^{-1+\frac{\varepsilon}{2}}(\tau)}{\Gamma(1-\frac{\varepsilon}{2})\varepsilon}
+\partial_0\int^\tau_{\tau_i}d\tau'\ a^{-1}(\tau')\log2\mu\Delta\tau\big\}\Big\}\notag\\
=&\frac{4i\pi^2\delta_\nu^{\ 0}\delta_\rho^{\ 0}}{(4-2\varepsilon)(2-2\varepsilon)}\Big\{
\frac{\pi^{-\frac{\varepsilon}{2}}\mu^{-\varepsilon}\partial_0\{a^{-1+\frac{\varepsilon}{2}}(\tau)\partial_0a^{-1+\frac{\varepsilon}{2}}(\tau)\}}{\Gamma(1-\frac{\varepsilon}{2})\varepsilon}\notag\\
&\hspace{8em}-H^2\log a(\tau)+2H^2+H^2\partial_0\big\{\tau\partial_0^2\int^\tau_{-1/\sqrt{p_\mu p^\mu}}d\tau'\ \tau'\log\frac{2\mu}{H}(1-\frac{\tau}{\tau'})\big\}\Big\}\notag\\
=&\frac{4i\pi^2H^2\delta_\nu^{\ 0}\delta_\rho^{\ 0}}{(4-2\varepsilon)(2-2\varepsilon)}\Big\{
\frac{\pi^{-\frac{\varepsilon}{2}}\mu^{-\varepsilon}(1-\frac{\epsilon}{2})(1-\epsilon)}{\Gamma(1-\frac{\varepsilon}{2})}\frac{a^\epsilon(\tau)}{\epsilon}-\log a(\tau)+2 \notag\\
&\hspace{8em}+\partial_0\big\{\tau\partial_0^2\int^\tau_{-1/\sqrt{p_\mu p^\mu}}d\tau'\ \tau'\log\frac{2\mu}{H}(1-\frac{\tau}{\tau'})\big\}\Big\}
\simeq 0\times\log a(\tau). \label{I21'}
\end{align}
As seen in the fourth line of (\ref{I21'}), the integral leads to an UV divergence but it is not related to the coefficient of $\log a(\tau)$. 
Specifically, time dependences are canceled between the first and second terms. 
That is a reasonable result because IR contributions and UV contributions can be clearly separated in comparison to Hubble scale. 
The integral of the fifth line corresponds to the non-local term. 
Here the lower bound of the integral has been corrected in a similar way to (\ref{bound}). 

We have found that 
(\ref{I22}) contributes to the coefficient of $\log a(\tau)$ but (\ref{I21}), (\ref{I23}) do not. 
It is consistent with their dependence under the scaling transformation. 
As a result, $I^2_{\ \mu\nu\rho\sigma\alpha\beta}$ is evaluated as 
\begin{align}
I^2_{\ \mu\nu\rho\sigma\alpha\beta}\simeq\frac{\kappa^2H^2}{4\pi^2}\log a(\tau)\times \eta_{\mu\beta}\eta_{\sigma\alpha}\delta_\nu^{\ 0}\delta_\rho^{\ 0}. 
\label{Is1.5}\end{align}
In the same way, we can confirm that 
only the dS broken term contributes to the coefficient of $\log a(\tau)$ in the other integrals of (\ref{Is1}). 
We list the results as follows
\begin{align}
I^1_{\ \mu\nu\rho\sigma\alpha\beta}&\simeq\frac{\kappa^2H^2}{4\pi^2}\log a(\tau)\times 
\eta_{\mu\beta}\{\eta_{\nu\alpha}\delta_\rho^{\ 0}\delta_\sigma^{\ 0}
+\eta_{\rho\alpha}\delta_\nu^{\ 0}\delta_\sigma^{\ 0}
+2\delta_\alpha^{\ 0}\delta_\sigma^{\ 0}\delta_\nu^{\ 0}\delta_\rho^{\ 0}\}, \label{Is2}\\
I^3_{\ \mu\nu\rho\sigma\alpha\beta}&\simeq\frac{\kappa^2H^2}{4\pi^2}\log a(\tau)\times 0, \notag\\
I^4_{\ \mu\nu\rho\sigma\alpha\beta}&\simeq\frac{\kappa^2H^2}{4\pi^2}\log a(\tau)\times 
\delta_{\mu}^{\ 0}\delta_\sigma^{\ 0}\{\frac{1}{2}\eta_{\nu\alpha}\eta_{\rho\beta}+\frac{1}{2}\eta_{\nu\beta}\eta_{\rho\alpha}
+4\delta_\nu^{\ 0}\delta_\rho^{\ 0}\delta_\alpha^{\ 0}\delta_\beta^{\ 0}\notag\\
&\hspace{8.5em}+\eta_{\nu\alpha}\delta_\rho^{\ 0}\delta_\beta^{\ 0}+\eta_{\nu\beta}\delta_\rho^{\ 0}\delta_\alpha^{\ 0}
+\eta_{\rho\alpha}\delta_\nu^{\ 0}\delta_\beta^{\ 0}+\eta_{\rho\beta}\delta_\nu^{\ 0}\delta_\alpha^{\ 0}
+\eta_{\alpha\beta}\delta_\nu^{\ 0}\delta_\rho^{\ 0}\},\notag\\
I^5_{\ \mu\nu\rho\sigma\alpha\beta}&\simeq\frac{\kappa^2H^2}{4\pi^2}\log a(\tau)\times
\eta_{\sigma\beta}\{\eta_{\nu\alpha}\delta_\mu^{\ 0}\delta_\rho^{\ 0}
+\eta_{\rho\alpha}\delta_\nu^{\ 0}\delta_\mu^{\ 0}
+2\delta_\mu^{\ 0}\delta_\alpha^{\ 0}\delta_\nu^{\ 0}\delta_\rho^{\ 0}\}. \notag
\end{align}
We can rederive (\ref{sint2}) by substituting (\ref{Is1.5}) and (\ref{Is2}) in (\ref{Is0}). 
It is reasonable since we can decompose the dS broken part of (\ref{minimallyD}) as follows
\begin{align}
-\frac{H^2}{8\pi^2}\log H^2\Delta x^2=-\frac{H^2}{8\pi^2}\log y+\frac{H^2}{8\pi^2}\log a(\tau)a(\tau'), 
\end{align}
and the dS invariant logarithm $\log y$ may not contribute to the coefficient of $\log a(\tau)$. 

\section{Local terms with IR logarithms from (\ref{self})}\label{A:C}
\setcounter{equation}{0}

In this appendix, we list the local contribution which comes from each term of the self interaction (\ref{self}). 
In a similar way to (\ref{Dint1}) and (\ref{Dexpansion}), the following local terms are extracted up to $\mathcal{O}(\log a(\tau))$ 
\begin{align}
&\int d^4x'\ \frac{\log \big(a(\tau)a(\tau')\big)}{H^2a(\tau)a(\tau')}\partial_\mu\partial^2\delta^{(4)}(x-x')
\hat{\psi}(x') \label{J1}\\
=&\ \partial'_\mu\partial'^2
\big\{\frac{\log \big(a(\tau)a(\tau')\big)}{H^2a(\tau)a(\tau')}\hat{\psi}(x')\big\}\big|_{x'=x} 
\simeq\frac{2\log a(\tau)}{H^2a(\tau)}\partial_\mu\partial^2\Big(\frac{\hat{\psi}(x)}{a(\tau)}\Big), \notag
\end{align}
\begin{align}
&\int d^4x'\ \frac{1}{H^2a(\tau)a(\tau')}\partial_\mu\partial^4
\big[\frac{\log \mu^2\Delta x^2_{++}}{\Delta x^2_{++}}-\frac{\log \mu^2\Delta x^2_{+-}}{\Delta x^2_{+-}}\big]\hat{\psi}(x') \label{J2}\\
\simeq&-\frac{2\log a(\tau)}{H^2a(\tau)}\partial_\mu\partial^4
\int d^4x'\ \big[\frac{1}{\Delta x^2_{++}}-\frac{1}{\Delta x^2_{+-}}\big]\frac{\hat{\psi}(x')}{a(\tau')} \notag\\
=&-4i\pi^2\times
\frac{2\log a(\tau)}{H^2a(\tau)}\partial_\mu\partial^2\Big(\frac{\hat{\psi}(x)}{a(\tau)}\Big), \notag
\end{align}
\begin{align}
&\int d^4x'\ \log \big(a(\tau)a(\tau')\big)\partial_\mu\delta^{(4)}(x-x')
\hat{\psi}(x') \label{J3}\\
=&\ \partial'_\mu\big\{\log \big(a(\tau)a(\tau')\big)\hat{\psi}(x')\big\}\big|_{x'=x} 
\simeq 2\log a(\tau)\partial_\mu\hat{\psi}(x), \notag
\end{align}
\begin{align}
&\int d^4x'\ \partial_\mu\partial^2
\big[\frac{\log \mu^2\Delta x^2_{++}}{\Delta x^2_{++}}-\frac{\log \mu^2\Delta x^2_{+-}}{\Delta x^2_{+-}}\big]\hat{\psi}(x') \label{J4}\\
\simeq&-2\log a(\tau)\partial_\mu\partial^2
\int d^4x'\ \big[\frac{1}{\Delta x^2_{++}}-\frac{1}{\Delta x^2_{+-}}\big]\hat{\psi}(x') \notag\\
=&-4i\pi^2\times2\log a(\tau)\partial_\mu\hat{\psi}(x), \notag
\end{align}
\begin{align}
&\int d^4x'\ \partial_\mu\partial_i^2
\big[\frac{\log \mu^2\Delta x^2_{++}}{\Delta x^2_{++}}-\frac{\log \mu^2\Delta x^2_{+-}}{\Delta x^2_{+-}}\big]\hat{\psi}(x') \label{J5}\\
\simeq&-2\log a(\tau)\partial_\mu\partial_i^2
\int d^4x'\ \big[\frac{1}{\Delta x^2_{++}}-\frac{1}{\Delta x^2_{+-}}\big]\hat{\psi}(x') \notag\\
=&\ 0\times \log a(\tau)\partial_\mu\hat{\psi}(x), \notag
\end{align}
\begin{align}
&\int d^4x'\ \partial_\mu\partial_i^2
\big[\frac{1}{\Delta x^2_{++}}-\frac{1}{\Delta x^2_{+-}}\big]\hat{\psi}(x') \label{J6}\\
=&\ 0\times\log a(\tau)\partial_\mu\hat{\psi}(x). \notag
\end{align}
Here we have used the identities 
\begin{align}
\partial^2\frac{1}{\Delta x_{++}^2}=4i\pi^2\delta^{(4)}(x-x'),\hspace{1em}\partial^2\frac{1}{\Delta x_{+-}^2}=0,  
\end{align}
\begin{align}
&\ \partial_\mu\partial_\nu\partial_\rho\int d^4x'\ 
\big[\frac{1}{\Delta x^2_{++}}-\frac{1}{\Delta x^2_{+-}}\big]\hat{\psi}(x') \\
=&-4i\pi^2\big\{\delta_\mu^{\ 0}\delta_\nu^{\ 0}\partial_\rho
+\delta_\mu^{\ 0}\delta_\rho^{\ 0}\partial_\nu+\delta_\nu^{\ 0}\delta_\rho^{\ 0}\partial_\mu
-2\delta_\mu^{\ 0}\delta_\nu^{\ 0}\delta_\rho^{\ 0}\partial_0\big\}\hat{\psi}(x). \notag
\end{align}
By summing up (\ref{J1})-(\ref{J6}), we can derive the effective equation of motion (\ref{WD}). 
In the self-energy (\ref{self}), the higher derivative term seems to be associated with the IR logarithm at first sight. 
Considering the fact that the non-local term does not contribute to the dS symmetry breaking, 
we have found that its coefficient is canceled between (\ref{J1}) and (\ref{J2}). 


\end{document}